\documentclass[12pt]{article}

\usepackage[english]{babel}

\usepackage{arxiv}
\usepackage[utf8]{inputenc} 
\usepackage[T1]{fontenc}    
\usepackage[hidelinks]{hyperref}
\usepackage{url}            
\usepackage{booktabs}       
\usepackage{amsfonts}       
\usepackage{nicefrac}       
\usepackage{microtype}      
\usepackage{lipsum}		
\usepackage{graphicx}
\usepackage{natbib}
\usepackage{doi}
\usepackage{amsmath,bm}
\usepackage{amsthm}
\usepackage{tikz}
\usepackage{amssymb}
\usepackage{algorithm}
\usepackage{algpseudocode}
\usepackage[most]{tcolorbox}
\usepackage{multirow}
\usepackage{mathrsfs}

\usetikzlibrary{calc,fit,positioning}
\usetikzlibrary{trees, matrix}
\usetikzlibrary{arrows.meta}
\graphicspath{ {./images/} }
\usetikzlibrary{patterns}
\usetikzlibrary{decorations.pathreplacing}
\usepackage{hf-tikz}
\usetikzlibrary{shapes,decorations,arrows,calc,arrows.meta,fit,positioning}

\tikzset{every matrix/.style={inner sep=-\pgflinewidth,
        matrix of math nodes, column sep=-\pgflinewidth,
        nodes={draw=black, font=\color{black}, minimum size=.75cm, anchor=center}}}
        
\tikzset{
    -Latex,auto,node distance =1 cm and 1 cm,semithick,
    state/.style ={ellipse, draw, minimum width = 0.7 cm},
    point/.style = {circle, draw, inner sep=0.04cm,fill,node contents={}},
    bidirected/.style={Latex-Latex,dashed},
    el/.style = {inner sep=2pt, align=left, sloped}
}

\newcommand{\Perp}{{\perp\!\!\!\perp}}

\newcommand{\btheta}{\boldsymbol{\theta}}
\newcommand{\bZ}{\boldsymbol{Z}} 
\newtheorem{theorem}{Theorem}

\newtheorem{Assumption}{Assumption}

\DeclareMathOperator{\E}{\mathbb{E}}
\DeclareMathOperator{\bX}{\boldsymbol{X}}
\DeclareMathOperator{\bx}{\boldsymbol{x}}
\DeclareMathOperator{\bU}{\boldsymbol{U}}

\newcommand\StateX{\Statex\hspace{\algorithmicindent}}

\algnewcommand{\Initialization}[1]{%
  \State \textbf{Initialization:}
  \Statex \hspace*{\algorithmicindent}\parbox[t]{.8\linewidth}{\raggedright #1}
}

\algnewcommand{\Input}[1]{%
  \State \textbf{Input:}
  \Statex \hspace*{\algorithmicindent}\parbox[t]{.8\linewidth}{\raggedright #1}
}

\algnewcommand{\Output}[1]{%
  \State \textbf{Output:}
  \Statex \hspace*{\algorithmicindent}\parbox[t]{.8\linewidth}{\raggedright #1}
}

\def\spacingset#1{\renewcommand{\baselinestretch}%
{#1}\small\normalsize} \spacingset{1}

\title{A Bayesian Approach to Estimate Causal Peer Influence Accounting for Latent Network Homophily}

\author{ 
    {Seungha Um} \\
	Division of Biostatistics\\
    Department of Population Health\\
	New York University Grossman School of Medicine\\
	New York, NY 10016\\
	\texttt{Seungha.Um@nyulangone.org} \\
\And
	{Tracy Sweet} \\
	Department of Human Development and \\Quantitative Methodology \\
	University of Maryland \\
    College Park, MD 20742\\
	\texttt{tsweet@umd.edu} \\
\And
	{Samrachana Adhikari} \\
	Division of Biostatistics\\
    Department of Population Health\\
	New York University Grossman School of Medicine\\
	New York, NY 10016\\
	\texttt{Samrachana.Adhikari@nyulangone.org} \\
}

\renewcommand{\shorttitle}{\textit{arXiv} Template}

\hypersetup{
pdftitle={Estimating peer effect},
pdfsubject={LSM},
pdfauthor={Seungha ~Um, Samrachana ~Adhikari},
pdfkeywords={peer effect, LSM},
}

\begin{document}
\maketitle
\spacingset{1.9} 
\begin{abstract}
Researchers have focused on understanding how individual's behavior is influenced by the behaviors of their peers in observational studies of social networks. Identifying and estimating causal peer influence, however, is challenging due to confounding by homophily, where people tend to connect with those who share similar characteristics with them. Moreover, since all the attributes driving homophily are generally not always observed and act as unobserved confounders, identifying and estimating causal peer influence becomes infeasible using standard causal identification assumptions. In this paper, we address this challenge by leveraging latent locations inferred from the network itself to disentangle homophily from causal peer influence, and we extend this approach to multiple networks by adopting a Bayesian hierarchical modeling framework. To accommodate the nonlinear dependency of peer influence on individual behavior, we employ a Bayesian nonparametric method, specifically Bayesian Additive Regression Trees (BART), and we propose a Bayesian framework that accounts for the uncertainty in inferring latent locations. We assess the operating characteristics of the estimator via simulation study. Finally, we apply our method to estimate causal peer influence in advice-seeking networks of teachers  in secondary schools and assess whether teachers' beliefs about mathematics education is influenced by the beliefs of their peers from whom they receive advice. Our results suggest that overlooking latent homophily can lead to either underestimation or overestimation of causal peer influence, accompanied by considerable estimation uncertainty.
\end{abstract}


\section{Introduction}

Social network data typically consist of a set of individuals and the relational ties among them. Within a social network, an individual's behavior may adapt or respond to the behaviors of others. This phenomenon of alters' behaviors impacting an ego's behavior is commonly referred to as peer influence. Identifying and quantifying peer influence has been a focus of research in a variety of disciplines, including substance use \citep{napier1984assessment, henneberger2021peer}, health outcomes \citep{christakis2007spread,Christakis2008}, and education \citep{supovitz2010principals, rosenqvist2018two}. As an example from education research, consider an advice-seeking network among teachers in a school.  In such a network, a teacher's individual beliefs about instruction could be influenced by the beliefs of their advisors and other teachers from whom they seek advice \citep{supovitz2010principals, Sweet2020}. Understanding peer influence in the network of teachers may be important for enhancing instructional methods and facilitating the dissemination of innovative teaching approaches. 

Despite the widespread interest in peer influence, identifying and accurately estimating a causal peer influence effect using observational network data is often challenging due to confounding by multiple measured and unmeasured factors that lead to homophily \citep{ogburn2018challenges}. Homophily is the tendency of individuals to associate and form connections with others who are similar to themselves \citep{mcpherson2001birds, ogburn2018challenges, Shalizi2011}. In our example with the advice-seeking network, two teachers in a school may seek advice from each other because they share similar beliefs, for example, as a consequence of both teaching the same grade. If they both also encourage their students to construct their own knowledge through exploration and experimentation, reflecting similar teaching beliefs, it is important to separate the influence of peer teaching beliefs from similarity in beliefs to establish causality.

Thus, to accurately identify causal peer influence, the effect of peers' behaviors should be distinguished from the effect of similarities, which may induce similar behaviors among peers \citep{Shalizi2011}. However, for many social networks it is not possible to observe or measure all homophilous attributes that lead to relational ties, resulting in latent homophily. Further, when there is confounding due to latent homophily, the peer influence effect is not causally identified without making additional assumptions \citep{Shalizi2011}. In fact, none of the previous work that focused on estimating peer influence in advice-seeking networks for teachers addressed confounding due to latent homophily \citep{Spillane2018, Sweet2020}. 

In this paper, we propose an approach to estimate causal peer influence effects while adjusting for confounding due to latent homophily. Specifically, we present identification assumptions for the causal peer influence effect in the presence of latent homophily, extending previous work  \citep{Macfowland2021, Cristali2022}, and develop a flexible Bayesian hierarchical framework for estimation. We further extend our proposed Bayesian hierarchical framework to estimate peer influence effect from multiple networks while accommodating the nonlinear dependency of peer influence on individual behavior. 






\subsection{Related Work} 



A common strategy to account for unobserved confounders in causal inference involves the use of proxies of unmeasured confounders \citep{Liu2024}. Proxy variables, commonly used in epidemiology studies and referred to as negative control variables, are employed to detect and sometimes correct for unmeasured confounding \citep{Tchetgen2014,Wang2018,Shi2020}. There are two primary approaches to leverage proxy variables: i) selecting a proxy variable from a set of plausible observed candidates, or ii) estimating a proxy variable as a latent variable using observed data. This strategy of using proxy variables is applicable regardless of whether individuals are independent or connected through network ties, with further elaboration provided below. 

The first approach of selecting a proxy variable leverages the association between an unmeasured confounder and a proxy variable, enabling researchers to establish the conditional distribution of the proxy variable given the unmeasured confounder. Recent papers employing such a proxy variable have focused on the identification and estimation of causal effects involving both non-network data \citep{Liu2024, Wang2018, Wang2023, Christos2017} and network data \citep{egami2021, Liu2022}. For example, \cite{Christos2017} viewed the observed proxy variables as noisy versions of unmeasured confounders and proposed a method that was robust against varying noise levels. In considering network data to account for latent homophily, \cite{egami2021} used two proxy variables; one was known not to be causally affected by the treatment of interest, and the other did not causally affect the outcome of interest.


However, identifying suitable proxy variables from observed candidates in real-world scenarios can often be challenging. To address this challenge, the second approach of estimating proxy variables may be more appropriate. In a network context, we can leverage information embedded within network ties to estimate an unmeasured homophilous attribute.  In fact, many existing latent variable social network models \citep{Dabbs2020} can be utilized to estimate such proxy variables. For example, consider a latent space network model \citep{Hoff2002} which assumes that each node in a network can be represented by a position in a lower dimensional latent space. Conditional on the latent positions, the probability of two nodes sharing a tie is inversely related to the distance between their latent positions in this space. Consequently, these latent positions, which capture the existing network structures including homophily, transitivity and clustering, can serve as a proxy for latent homophilous attributes. Adjusting for these inferred latent positions, \cite{Macfowland2021} introduced a consistent estimator of causal peer effects. Alternatively, \cite{Cristali2022} used embedding methods to capture the latent properties of each node as proxies and provided a nonparametric identification of causal peer influence. However, both studies are limited to investigations assessing a linear relationship between ego's  and peers' behaviors, where a coefficient in a linear model was interpreted as the causal peer influence effect \citep{Macfowland2021,Cristali2022}.  Such restrictive assumptions of linearity presents challenges for modeling nonlinear relationships between behavior and peer influence, including interactions between predictors. Also, both of these studies on the estimation of causal peer influence are examined under a single network setting. Therefore, a comprehensive and flexible causal framework for assessing peer influence in multiple social networks is lacking.

\normalcolor

\subsection{Motivating application}\label{subsec:mot}

\begin{table}[t]
\centering
\begin{tabular}{lllllllll}
  \hline
 ID & size & degree & density & transitivity & score (2012) & peers' score (2012) & score (2013)  & difference \\ 
  \hline
  S1 &  27 & 3.1(1.4) & 0.2 & 0.34 & 19.8(2.6) & 19.8(1.3) & 19.9(2.5) & 0.9(0.7) \\ 
  S2 &  27 & 3.6(3.0) & 0.3 & 0.32 & 18.4(2.2) & 18.5(3.8) & 19.4(1.6) & 1.8(1.5) \\ 
  S3 &  25 & 3.0(1.1) & 0.3 &  0.42 & 19.0(2.4) & 19.1(1.4) & 19.5(2.6) & 1.6(1.6) \\ 
  S4 &  30 & 3.5(1.3) & 0.2 & 0.51& 19.5(2.1) & 19.1(3.8) & 20.2(2.3) & 1.5(1.0) \\ 
  S5 &  23 & 2.2(1.3) & 0.2 & 0.31 & 19.3(2.4) & 18.9(4.6) & 19.2(1.8) & 1.8(1.5) \\ 
  S6 &  26 & 2.3(1.0) & 0.2 & 0.46& 17.8(2.3) & 17.1(3.8) & 18.6(2.2) & 1.7(1.4) \\ 
  S7 &  20 & 2.3(1.7) & 0.2 & 0.37 & 17.7(2.3) & 16.0(5.7) & 19.0(1.7) & 2.0(1.7) \\ 
  S8 &  26 & 3.4(1.5) & 0.3 &0.45& 18.8(2.1) & 18.7(1.2) & 19.7(1.7) & 1.5(1.3) \\ 
  S9 &  24 & 2.6(1.1) & 0.2 & 0.14 &18.2(2.7) & 18.1(2.1) & 16.9(1.4) & 2.2(2.2) \\ 
  S10 &  28 & 2.8(1.3) & 0.2 & 0.39 & 19.1(1.8) & 18.3(4.0) & 18.0(1.6) & 1.9(1.9) \\ 
  S11 &  23 & 2.2(1.1) & 0.2 & 0.21&19.1(2.6) & 17.8(4.4) & 18.9(2.2) & 1.8(1.8) \\ 
  S12 &  22 & 2.2(0.7) & 0.2 & 0.55& 18.7(2.3) & 18.8(1.3) & 19.6(2.2) & 2.3(2.4) \\ 
  S13 &  15 & 2.6(1.9) & 0.4 & 0.18 & 18.8(2.4) & 18.1(5.3) & 19.1(2.1) & 1.9(1.3) \\ 
  S14 &  13 & 2.5(1.6) & 0.4 & 0.16& 19.8(2.0) & 17.3(7.8) & 19.8(2.4) & 1.4(1.3) \\ 
   \hline
\end{tabular}
\caption{Summary statistics of the network characteristics and teaching belief scores for each school with mean (sd). The difference represents the absolute difference between the scores from 2012 and 2013.}\label{tab:summary}
\end{table}

Returning to our example of advice-seeking networks, our work is motivated by a real-world scenario involving elementary school staff in a mid-sized Midwestern school district referred to as Auburn Park \citep{Pitts2009}. In 2012, this suburban district served approximately 5,800 students across 14 elementary schools. In Spring 2012 and Spring 2013, staff members in each school responded to surveys that included items related to their beliefs about mathematics education and about to whom they sought advice and information about curriculum and instruction. In their study on mathematics beliefs, \cite{Spillane2018} identified a mathematics-related belief construct that they called student-centered learning; this was based on 5 Likert scale items that advocated for engaging students with guiding questions rather than direct instruction, promoting an environment where students can construct their own knowledge through exploration. They used the sum score to generate an overall belief score for each respondent, which we will use as the outcome of interest in our study. Details on the specific survey items can be found in the Appendix (Table \ref{tab:survey}) and summary statistics of the network characteristics and teaching belief scores for each school are represented in Table \ref{tab:summary}. In Table \ref{tab:summary}, for each school we present the size of the network (size), average degree (average number of ties per node), network density (the ratio between the actual number of ties and the maximum possible number of ties), transitivity (the ratio between the observed number of closed triplets and the maximum possible number of closed triplets), along with a summary of the belief scores in 2012 and 2013. The advice-seeking networks observed across the 14 schools varied in their size, ranging from 13 to 30 teachers, and also in their network structures with network densities ranging from 0.18 to 0.41. Further information about the dataset is available in \cite{Spillane2018} and \cite{Sweet2020}.

\begin{figure}[!t]
    \centering
    \includegraphics[width=1 \linewidth]{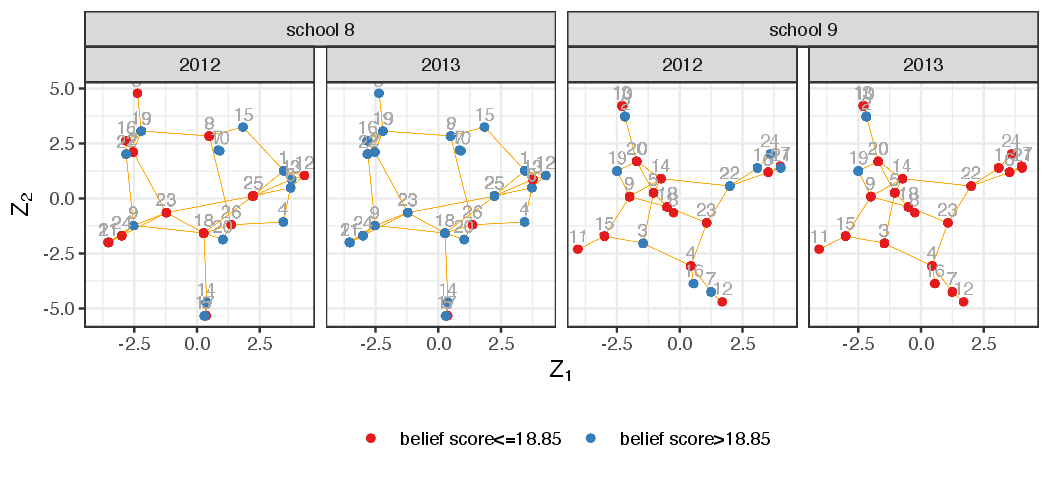}\vspace{-0.8cm}
    \caption{The teachers' scores about student-centered beliefs in 2012 and 2013 for school 8 and school 9 with network ties. The belief scores are represented visually by coloring each node based on high/low belief scores, using a cut-off of $18.85$ for effective visualization.}
    \label{fig:net_real_obs}
\end{figure}

We can visualize changes in teachers' individual beliefs from 2012 to 2013, potentially in response to their peers' belief scores in 2012. As an illustration, Figure \ref{fig:net_real_obs} shows the belief scores for advice-seeking networks in School 8 and School 9; similar plots for the remaining schools are shown in Figure \ref{fig:net_real_obs_all} in the Appendix. For effective visualization, we dichotomized the continuous math belief score into high and low categories using a cut-off of $18.85$, the average belief score from 2012. Red nodes are nodes with low belief scores, and blue nodes are those with high belief scores.

As shown in Figure \ref{fig:net_real_obs} (left), 17 teachers (65.3\%) in School 8 exhibit high belief scores in 2012, and 23 teachers (88.4\%) have high scores in 2013. This suggests a plausible shift towards higher scores in 2013. Conversely, for School 9 (right), 15 teachers (57.5\%) had low scores in 2012, while 22 teachers (84.6\%) had low scores perhaps in 2013. These observations suggest that individuals tended to have lower scores in 2013 when their peers had low scores in 2012, and higher scores when their peers had high scores, which suggests some evidence of peer influence. 



Our primary goal in this paper is to assess whether the teachers in each school changed their beliefs about mathematics education based on the beliefs of their peers from whom they receive advice. In order to establish that the belief scores of the teachers in 2013 were causally influenced by their peers scores in 2012, we need to rule out all other common causes. However, this assumption is likely to be violated in advice-seeking networks because teachers often seek advice from peers who share similar attributes and beliefs, which can also contribute to changes in their belief scores. Many of these attributes are, however, not fully observed. Estimates of causal peer influence effects that do not adjust for these unobserved homophilous attributes are likely going to be biased. Therefore, to isolate the influence of peers from the effects of these unobserved attributes, we propose a Bayesian approach to estimate and adjust the latent homophilous variable. 


\subsection{Contribution}

This paper presents a novel Bayesian framework for estimating causal peer influence while incorporating latent variables to address unmeasured confounding due to homophily with a semiparametric approach. The framework is designed for scenarios where independent yet similar networks are observed simultaneously, and there is a nonlinear relationship between individual behavior and peers' behaviors. Our main contributions are as follows.

First, we address the issue of disentangling homophily from peer influence by leveraging latent locations inferred from the network itself, instead of using observed proxy variables. This extends the work of \cite{Macfowland2021} to integrate the estimation of peer influence within a causal framework by clearly outlining causal identification assumptions with latent homophily. Furthermore, our work focuses on inferring latent locations from the continuous latent space model which can be viewed as a generalized version of the latent community model considered in simulation studies by \citet{Macfowland2021} .

Second, in contrast to the existing literature, which has largely focused on estimating causal peer influence within a single network \citep{Macfowland2021,Cristali2022,egami2021}, we extend the latent locations strategy to multiple networks through a Bayesian hierarchical modeling framework \citep{Sweet2020}. This framework facilitates the estimation of shared homophilous attributes, represented as latent locations, across different networks. Additionally, it allows us to quantify the uncertainty in inferring these latent positions, which is particularly useful when each observed network has a small number of nodes.

Lastly, we adopt BART, a Bayesian nonparametric model, to address the limitations of linear models in capturing complex, non-linear relationships between individual behavior and peers' behaviors while also facilitating uncertainty quantification using posterior predictive distributions. Existing research has primarily focused on estimating causal inference using linear regression models, often overlooking the aspect of uncertainty quantification \citep{Macfowland2021,egami2021,Forastiere2021}. Although \cite{Forastiere2022JMLR} employed penalized splines to accommodate nonlinear dependencies between outcomes and neighborhood treatments without homophily, BART efficiently captures nonlinearity and offers the additional flexibility of not requiring a predefined functional form. To the best of our knowledge, this is the first attempt to incorporate the estimation of homophily attributes within a Bayesian framework for identifying causal peer influence and extend it to the multiple networks setting.

The remainder of this article is organized as follows. In Section \ref{sec:notation}, we introduce notation and establish the nonparametric identification of causal peer influence, outlining the necessary conditions. In Section \ref{sec:model}, we propose estimation strategies for the causal peer influence effect within a Bayesian framework. We then evaluate our methods through a simulated study in Section \ref{sec:simulation}, and we apply our proposed methodology to the motivating advice-seeking network dataset in Section \ref{sec:real}. Section \ref{sec:discussion} concludes with a discussion.

\color{black}

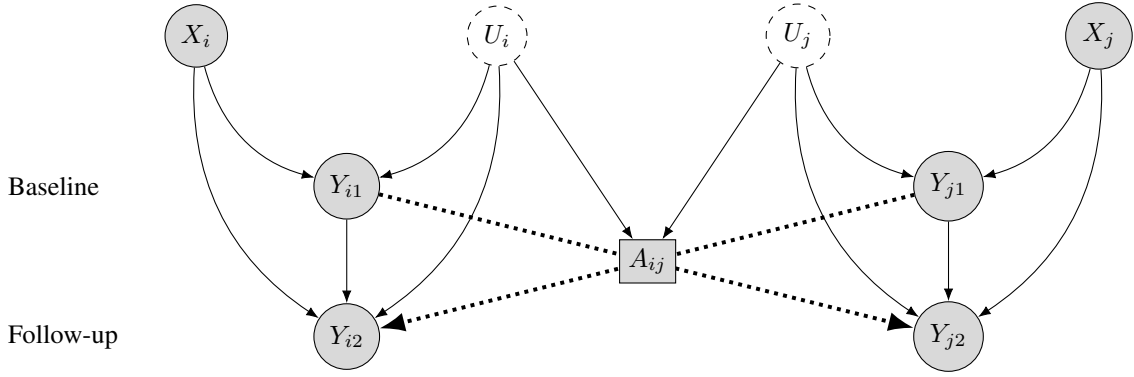
\begin{figure}
    \centering
        \begin{tikzpicture}
            \node[text width=3cm] at (-1, -2) (P1) {Baseline};
            \node[text width=3cm] at (-1, -4) (P1) {Follow-up};
            \node[state, circle, fill = gray!30] (1) at (0, 0) {$X_i$};
            \node[state, dashed, circle] (2) at (4, 0) {$U_i$};
            \node[state, dashed, circle] (3) at (8, 0) {$U_j$};
            \node[state, circle, fill = gray!30] (4) at (12, 0) {$X_j$};
            \node[state, circle, fill = gray!30] (5) at (2, -2) {$Y_{i1}$};
            \node[state, circle, fill = gray!30] (6) at (2, -4) {$Y_{i2}$};
            \node[state, circle, fill = gray!30] (7) at (10, -2) {$Y_{j1}$};
             \node[state, circle, fill = gray!30] (8) at (10, -4) {$Y_{j2}$};
            \path (5) edge (6);
            \path (7) edge (8);
            \path (1) edge [bend right] (5);
            \path (1) edge [bend right] (6);
            \path (2) edge [bend left] (5);
            \path (2) edge [bend left] (6);
            \path (3) edge [bend right] (7);
            \path (3) edge [bend right] (8);
            \path (4) edge [bend left] (7);
            \path (4) edge [bend left] (8);
            \path (5) edge [line width=1.5pt, dotted]  (8);
            \path (7) edge [line width=1.5pt, dotted] (6);
            \node[state, rectangle, fill = gray!30] (9) at (6, -3) {$A_{ij}$};
            \path (2) edge (9);
            \path (3) edge (9);
        \end{tikzpicture}
    \caption{The graphical causal model for our setting. The shaded (unshaded) ones represent observed (unobserved) variables. The solid lines indicate causal relations, while dotted lines indicate the peer effect of interest. For simplicity, we illustrate causal relationship between two connected nodes.}
    \label{fig:DAG}
\end{figure}

\section{Causal Inference in the Presence of Latent Homophily}\label{sec:notation}

\subsection{Notation}

We denote a network as $\mathcal{A}_n=(V_n,A)$, where $V_n$ is a set of nodes with cardinality $n$ and $A$ is a $n\times n $ adjacency matrix such that $A_{ij}=1$ represents the presence of a link from node $i$ to node $j$ for $i \neq j \in V_n$, and $A_{ij}=0$ otherwise. A network is undirected if $A^T = A$; otherwise the network is directed. We use $\mathcal{N}_i$ to denote a set of neighbors that share a tie with node $i$. We assume that a network is fixed over time and the behavior of node $i$ is observed at baseline $(t=1)$ and a single follow-up $(t=2)$. The behavior of node $i\in\{1,\cdots,n\}$ at time $t\in\{1,2\}$ is denoted by $Y_{it}$ for node $i$. To measure the peer influence on the behavior of node $i$ at follow-up ($Y_{i2}$), it is possible to consider a multivariate variable of peers' behaviors, $\{ \boldsymbol{Y}_{j1} : j\in \mathcal{N}_i\}$ for node $i$. However, in practice, the peer influence is assumed to be explained through a specific function $\phi(\cdot)$ designed for dimension reduction, known as an exposure mapping \citep{Aronow2017}. We define the summary of peers' behaviors at baseline for node $i$ as $G_i = \phi(\boldsymbol{Y}_{j1}: j\in \mathcal{N}_i)\in\mathbb{R}$. A common choice for $G_i$ is the average of peers' behaviors at $t=1$, that is, $G_i= \sum_{j \in \mathcal{N}_i} Y_{j1} / \sum_{j}A_{ij}$ \citep{Forastiere2021,egami2021}. Alternatively, we can consider the weighted mean of peers' behaviors as $G_i=\sum_{j \in \mathcal{N}_i} w_{ij}A_{j1}/\sum_{j\in\mathcal{N}_i}w_{ij}$, where each peer has a distinct level of impact on the behavior of node $i$. Throughout the paper, we consider the average of peers' behaviors to define $G_i$ for node $i$.


Let $\bX_i$ denote a vector of observed covariates for node $i$, and $U_i$ denote an unmeasured confounder that affects not only behaviors at baseline ($Y_{i1}$) and follow-up ($Y_{i2}$), but also the network ties $A_{ij}$ for $j\in \mathcal{N}_i$. The corresponding directed acyclic graph (DAG) is represented in Figure \ref{fig:DAG}. To estimate the causal peer influence of node $j$ on node $i$ using observational data, we condition on the network tie $A_{ij} = 1$ (i.e. presence of network ties). However, when conditioning on the collider $A_{ij} = 1$ , there is an unblocked backdoor path $Y_{j1} \leftarrow U_j \rightarrow A_{ij} \leftarrow U_i \rightarrow Y_{i2}$ where $U_j$ and $U_i$ are unmeasured \citep{Shalizi2011, egami2021}. Therefore, in the presence of unmeasured confounder $U_i$, identification and estimation of the causal peer influence is not feasible without additional assumptions. We outline the identification assumptions that rely on the latent homophily property of a network in Section \ref{subsec::identification}. More specifically, we leverage on the observation that when a network is formed based on homophily, it is likely that a node will be similar to its neighbors, indicating that the values of $U_i$ and $U_j$ are close. 

\subsection{Causal Estimand: Causal Peer Influence}
Using the potential outcomes framework \citep{Neyman1990,rubin1974}, we define $Y_{i2}(G_i = g)$ as the potential outcome for node $i$'s behavior at follow-up. The potential outcome, simplified as $Y_{i2}(g)$, denotes the behavior of node $i$  that would have been observed at follow-up if the average of peers' behaviors at baseline were $g$. To measure the influence of peer behavior on individual's behavior, 
we can define the average causal peer influence (ACPI) given the observed network $\mathcal{A}_n$, denoted by $\tau(g,g';\mathcal{A}_n)$, using the potential outcome notation as 
\begin{align*}
    \tau(g, g';\mathcal{A}_n) = E \left[Y_{i2}(g) -  Y_{i2}(g') |i\in\mathcal{A}_n \right]= \frac{1}{n}\sum_{i\in \mathcal{A}_n}\left[Y_{i2}(g) -  Y_{i2}(g')  \right] =\mu(g) -\mu(g')
\end{align*}
\normalcolor
for $g,g' \in\mathcal{G}$ where $\mathcal{G}$ is the support of $G_i$. The causal estimand is written as the empirical mean \citep{Forastiere2021,Forastiere2022JMLR,egami2021}, which can be viewed from a "super-population" perspective in terms of expectation \citep{Imbens2015,Forastiere2021}. The expectation is conditional on the observed network $\mathcal{A}_n$ and we write $\tau(g, g';\mathcal{A}_n)=\tau(g, g')$ omitting dependency on $\mathcal{A}_n$ for notational simplicity. 

This estimand compares the average potential outcomes at two distinct hypothetical peers' behaviors $g$ and $g'$, while preserving the identical set of individuals connected by the same network ties. The ACPI depends on the quantity of the average potential outcome at level $g$, denoted by $\mu(g) = E(Y_{i2}(g)|\mathcal{A}_n)$, which captures the causal relationship between peers' and individual behavior as $g$ varies in $\mathcal{G}$. We can also view $\mu(g)$ as an average dose-response function that depends on the dose of treatment for the subset $\mathcal{A}_n$ \citep{Forastiere2021}. In the presence of network dependency, causal estimands are commonly defined by conditioning on the specific network, $\mathcal{A}_n$, without specifying the sampling mechanism from a larger population. This is because describing the sampling mechanism is challenging to conceptualize in network settings \citep{Forastiere2022JMLR}, and the average of the potential outcome converges to a fixed quantity independent of the particular network $\mathcal{A}_n$ under some conditions \citep{Cristali2022}. It is worth noting that the potential outcome can be defined based on both $Y_{i1}$ and $G_i$, as in $Y_{i2}(Y_{i1}=y, G_i=g)$. However, when the goal is to estimate peer influence while maintaining the observed behaviors of node $i$ and varying only the behaviors of peers, the potential outcome can be defined solely depending on $g$ \citep{egami2021}. 


\subsection{Identifying Assumptions}\label{subsec::identification}
Given that the ACPI involves comparing two distinct quantities of potential outcome at different levels of $g$, the identification of the average potential outcome allows us to identify the ACPI. We outline the necessary assumptions required for the identification of the average potential outcome as following.

\begin{Assumption}\label{assump:consistency} (Consistency) 

Given a function of $\phi:\mathbb{R}^{|\mathcal{N}_i|}\rightarrow\mathbb{R}, \ \forall i\in \mathcal{A}_n$, 
\begin{align}
Y_{i2} = Y_{i2}\left(g\right) \text{ if } g = \phi(\boldsymbol{Y}_{j1}:j\in\mathcal{N}_i). 
\end{align}
\end{Assumption}

Assumption \ref{assump:consistency} implies that an individual's potential outcome under the observed peers’ behaviors at baseline is the outcome that will actually be observed for that person.

\begin{Assumption}\label{assump:Nei_interference} (Neighborhood interference) 

If $\phi(\boldsymbol{Y}_{j1}:j\in\mathcal{N}_i) = \phi(\boldsymbol{Y}'_{j1}:j\in\mathcal{N}_i)$ for all $i\in \mathcal{A}_n$, then
\begin{align} 
Y_{i2}\left( \phi(\boldsymbol{Y}_{j1})\right) = Y_{i2}\left(\phi(\boldsymbol{Y}^\prime_{j1})\right).
\end{align}
\end{Assumption}

Assumption \ref{assump:Nei_interference} states that the potential outcomes take the same value when the function values of different combinations of the peers' behavior are equal. Assumptions \ref{assump:consistency} and \ref{assump:Nei_interference} can be interpreted as stable unit treatment on neighborhood value assumption (SUTVA) in the presence of network interference \citep{Forastiere2021}.


\begin{Assumption}\label{assump:positivity} (Positivity)
\begin{align}
0 < P(G_i = g|Y_{i1}=y, \bX_i=x, U_i=u) <1 \quad \text{for all } g,u,x, \text{ and }y
\end{align}
\end{Assumption}

This assumption states that each individual has some chance of having a possible value of $g\in\mathcal{G}$. 
However, this assumption of positivity may not always be satisfied. For example, isolated nodes without any network connection do not meet this assumption because the function values of $g$ cannot be calculated ($\mathcal{N}_i = \emptyset$). Isolated nodes could be treated and analyzed separately or considered as $G_i = 0$, as if they have peers and their peers' behaviors were 0 \citep{Forastiere2021}. This scenario is beyond the scope of our paper.


\begin{Assumption}\label{assump:ignorability} (Latent unconfoundedness) 
\begin{align}
Y_{i2}(g) \Perp G_i | Y_{i1}, U_i, \bX_i \quad  \forall i, \ g\in \mathcal{G}.    
\end{align}
\end{Assumption}

Assumption \ref{assump:ignorability} implies that one's potential outcome is independent of  peers' behaviors at baseline conditional on the individual behavior $Y_{i1}$, observed confounder $\bX_i$ and unmeasured confounder $U_i$. Therefore this assumptions suggests that conditioning on $U_i$ is needed to control for confounding. 
However, as noted in \cite{egami2021}, Assumption \ref{assump:ignorability} while necessary, is not sufficient for potential outcome identification because we do not observe $U_i$.



\begin{Assumption} (Proxy variable for homophilous attributes)

There exists a proxy $Z_i$ for the unmeasured confounder $U_i$, and $U_i$ is $Z_i$-measurable, satisfying the following condition:

 \begin{align}\label{assump:proxy}
     Y_{i2} \perp A_{ij} | Z_i, G_i, Y_{i1}, \bX_i.
 \end{align}
\end{Assumption}

The Assumption \ref{assump:proxy} indicates that conditioning on $Z_i$ removes the effect of conditioning on the $A_{ij}$, thereby preventing  the opening of the backdoor path. The measurability assumption implies that $U_i$ can be determined using the information provided by $Z_i$ \citep{Cristali2022}.





\begin{theorem}\label{thm:identification}
Under Assumptions \ref{assump:consistency}-\ref{assump:proxy}, we have 
\begin{align}\label{theorem}
    \E \left[Y_{i2}(g)|\mathcal{A}_n\right] &= \E \left[Y_{i2} | Y_{i1} = y, G_i = g, \bX_i = \bx, Z_i = z\right]\\
    &= \E \left[Y_{i2} | Y_{i1} = y, G_i = g, \bX_i = \bx, Z_i^* = z^*\right]
\end{align}
\end{theorem} 

where $Z_i^* = WZ_i$ and $W$ is an orthogonal matrix satisfying $W^\prime W=I$. Therefore, the average potential outcome can be identified by leveraging a proxy variable $Z_i$ or a transformed proxy variable $Z^{*}$, the observed data ($y$, $g$, $x$), and a set of nodes $i \in \mathcal{A}_n$. The proof is provided in the Appendix. 

The identifying assumptions we presented here, excluding the assumption on the proxy variable $Z_i$ (Assumption \ref{assump:proxy}), are similar to those in \cite{egami2021} and \cite{Liu2022}, where an observed variable associated with $U_i$ is selected for $Z_i$. However, when $Z_i$ is unobserved, as it is in our case, we can regard a proxy variable $Z_i$ as a latent variable that serves as a close approximation of $U_i$. This proxy variable can be learned by applying a latent variable social network model to estimate the latent structure that captures the homophily \citep{Christos2017}. 
Therefore, we further assume that we have access to such model which allows us to sample from the posterior distribution $f(Z_i|\mathcal{A}_n)$ conditional on the observed network $\mathcal{A}_n$.



There is an inherent trade-off between the bias that results from omitting the unmeasured homophilous confounder $U_i$ and the bias from imprecisely estimating or selecting the proxy variable $Z_i$. However, if the proxy variable $Z_i$ contains sufficient information about $U_i$, accounting for $Z_i$ can substantially mitigate bias compared to omitting it. Additionally, as covariates measured in an observational study are not always eligible as proxy variables for the unmeasured homophilous confounder, our emphasis in this paper is on estimating $Z_i$ from the observed network. We leverage existing latent variable network models \citep{Dabbs2020} to estimate the proxy variable that represents the underlying latent homophily in the network. This estimated proxy variable enables the identification in Theorem \ref{thm:identification} satisfying the assumptions presented in Section \ref{subsec::identification}. In this paper, we focus on a continuous latent variable $Z_i$, however the method can be extended to accommodate discrete latent variables, such as block membership in a stochastic block model \citep{Fortunato2010,HOLLAND1983,Karrer2011}.  




\section{Hierarchical Bayesian Model for the Estimation of Average Causal Peer Influence Effects}\label{sec:model}

In this section, we introduce estimation strategies for the causal estimand, ACPI, within a Bayesian framework. To estimate the ACPI, we rely on the posterior samples from $f(Z_i|\mathcal{A}_n)$ utilizing a well established latent variable network model, a latent space model with a binary tie value \citep{Hoff2005}. We also demonstrate the eligibility of the resulting estimates of latent space positions as proxy variables satisfying the required conditions, and we extend our strategy to accommodate multiple network settings. Further, after conditioning on the estimated $Z_i$, we incorporate a Bayesian nonparametric model, for the outcome model. We select a Bayesian additive regression tree (BART) for its ease-of-use and excellent predictive performance \citep{Chipman2010}.

\subsection{Latent Space Network Model}

We consider the latent space model (LSM) \citep{Hoff2002} which assumes that the formation of network tie $A_{ij} $ from node $i$ to $j$ depends on similarity between latent variables $Z_i$ and $Z_j$. In this setting, the latent variable $Z_i$, is the position of node $i$ in a continuous metric space (often the Euclidean metric), and the probability of forming a tie $p(A_{ij}=1)$ is inversely related to the distance between $Z_i$ and $Z_j$. Conditioning on these latent positions, which account for the existing network structure such as homophily, transitivity and clustering, network ties are assumed to be independent. The LSM can be written as 
\begin{align}\label{eq:lsm}
&A_{i j} \stackrel{iid}{\sim} \operatorname{Bernoulli}\left(p_{i j}\right), \quad \text {for } i,j=1,\cdots,n \\
&\operatorname{logit}\left(p_{i j}\right)= \alpha_0 - \beta_z||Z_i - Z_j||_2\nonumber
\end{align}

where $Z_i\in \mathbb{R}^d$ is the position of individual $i$ in the latent space. Denote $\boldsymbol{Z}=(Z_1,\cdots, Z_n)$. The intercept $\alpha_0$ represents a baseline probability of a tie for nodes in the same position and $\beta_z$ is the corresponding slope coefficient for the distances between $Z_i$ and $Z_j$. To estimate latent locations and parameters, both likelihood-based inference and Bayesian inference are feasible approaches \citep{Hoff2002}. In this paper, we will estimate LSM parameters within a Bayesian framework, which allows for uncertainty quantification and is useful for inference with small sample sizes \citep{Lee2004}.

The prior distributions are generally specified as:
\begin{align}\label{LSM:prior}
&Z_i, Z_j \sim MVN_d (0, \sigma_z I)\\
&\alpha_0 \sim N(\mu_\alpha, \sigma_\alpha^2)\nonumber\\
&\beta_z\sim N(\mu_\beta,\sigma_\beta^2)\nonumber
\end{align}

where $Z_i$ and $Z_j$ are independently sampled from a multivariate normal distribution with covariance matrix $\sigma_z I$ where $I$ is the $d$-dimensional identity matrix.
The intercept $\alpha_0$ and coefficient $\beta_z$
are assigned normal prior distributions with hyperparameters $(\mu_\beta,\sigma_\beta^2)$ and $(\mu_\alpha, \sigma_\alpha^2)$ for mean and variance, respectively.

The LSM can further be extended to model multiple networks using a hierarchical latent space model (HLSM; \cite{Sweet2013}). By introducing a network superscript $k$ to the variables in Equation \eqref{eq:lsm}, we define $A_{ij}^k$ as the value of the tie from individual $i$ to individual $j$ in network $k$. We denote the $k$-th adjacency matrix by $A^k$, where $k=1, \cdots, K$, and $\boldsymbol{Z}^k = (Z_1^k, \cdots, Z_{n_k}^{k})$ as the latent positions for network $k$ with $n_k$ nodes. Following \cite{Sweet2013}, the HLSM can be written as 
\begin{align}\label{eq:HLSM}
& A_{i j}^k \sim \operatorname{Bernoulli}\left(p_{i j}^k\right) \\
& \operatorname{logit}\left(p_{i j}^k\right)=\alpha_0^k - \beta_z^\prime\left\|Z_{i}^k-Z_{j}^k\right\|. \nonumber
\end{align}
The ties are generated from the model based on each $\alpha_0^{k}$ and $Z_i^k$, determining the structure of the k$^{th}$ network. To effectively model networks that are independent yet similar, the following hierarchical structure is considered:

\begin{align}
& Z_{i}^k, Z_{j}^k \stackrel{i i d}{\sim} M V N_d(0, \sigma_z' I)  \\
&\alpha_0^k \sim N(\mu_{\alpha'}, \sigma_{\alpha'}^2) \nonumber\\
&\beta_z^\prime\sim N(\mu_{\beta'},\sigma_{\beta'}^2).\nonumber
\end{align}

For $k=1, \ldots, K$, the parameters $\alpha_0^{k}$ and $\beta_z^\prime$ as well as network specific latent locations $Z_i^k$ are independently drawn from a common population. This common population enables the effective modeling of networks that are independent yet similar.


Both the LSM and the HLSM model tie formation as a function of the distance between nodes. The distances between a collection of nodes in Euclidean space are preserved under rotation, reflection, and translation \citep{Hoff2002}; this implies that an infinite number of node positions results in the same log-likelihood, given that $\log \Pr(A|\boldsymbol{Z},\alpha_0,\beta_z) = \log\Pr(A|\boldsymbol{Z}^*,\alpha_0,\beta_z)$ for any $\boldsymbol{Z}^*$ equivalent to $\boldsymbol{Z}$ through the transformation where $\boldsymbol{Z}^* = (Z_1^*,\cdots,Z_n^*)$. Consequently, the latent positions $\boldsymbol{Z}$ are identifiable only up to the class of distance preserving transformations. However, this non-identifiability can be addressed by using a Procrustes transformation \citep{borg2005}, to align each of the latent position posterior draws with a fixed target position $\boldsymbol{Z}_0$ as a post-processing step. The Procrustes transformation can be implemented by targeting $\bZ^*=\arg \min _{W \bZ} \operatorname{tr}\left(\bZ_0-W \bZ \right)^{\prime}\left(\bZ_0 - W \bZ\right)$, where $W$ is an orthogonal matrix $W \in\mathbb{R}^{d\times d}$ with $W^TW=I$. This can easily be derived by $\bZ^* = \bZ_0\bZ'(\bZ \bZ_0' \bZ_0 \bZ')^{-1/2}\bZ$ assuming $\bZ_0$ and $\bZ$ are both centered at the origin. The fixed target positions $\boldsymbol{Z}_0$ can be set as the initial latent positions in the MCMC algorithm by employing multidimensional scaling (MDS) on the observed distance matrix. For the detail, we refer to \cite{Hoff2002,Adhikari2021}. 

In an HLSM, the Procrustes transformation is performed on the posterior draws of the latent positions $\boldsymbol{Z}^K$ for each network $k$ where each network has a different fixed target positions $Z_0^k$ such that $\bZ^{*k}=\arg \min _{W^k \bZ^k} \operatorname{tr}\left(\bZ^k_0-W^k \bZ^k \right)^{\prime}\left(\bZ^k_0 - W^k \bZ^k\right)$ for $k=1,\cdots, K$. This is particularly important when $n_k \neq n_{k'}$ for $k,k' \in (1,\cdots,K)$ and  $k\neq k'$. In the multiple network setting, $\bU^k$ is identifiable as $W^k Z^k$ and the rotation matrix $W^k$ varies across networks.

\subsection{Outcome Model with Bayesian Additive Regression Trees}

Let $\mathcal{D}_i = (Y_{i1}, G_i , Z_i, \bX_i)$ denote a vector of predictors for node $i$,  where $Z_i$ represents the latent location sampled from the LSM. Then the outcome model for behavior at follow-up, $Y_{i2}$ for node $i\in (1,\cdots,n)$ is written as
$$
Y_{i2}= f(\mathcal{D}_i)+\epsilon_i
$$
where $\epsilon_i \sim \mathrm{N}(0,\tau^2)$. To model the unknown mean function $f(\cdot)$, we consider BART, which offers a flexible and nonparametric approach to model multidimensional regression bases being learned directly from the observed data\citep{Chipman2010}. Such a nonparameteric approach accommdates a variety of nonlinear associations between peer influence and behavior.

The BART framework introduced by  models $f(\mathcal{D}_i)$ as a sum of $m$ trees
\begin{align}
  f(\mathcal{D}_i)= \sum_{t=1}^m g(\mathcal{D}_i; \mathcal{T}_t, \mathcal{M}_t)\ ,\label{eq:BART}
\end{align}

where $\mathcal{T}_t$ is a binary tree structure with $n_t$ leaf nodes and $\mathcal{M}_t = \{\mu_{t1}, \ldots, \mu_{tn_t}\}$ is a collection of parameters for $n_t$ leaves of the $t$th tree $\mathcal{T}_t$. The function $g(\mathcal{D}_i; \mathcal{T}_t, \mathcal{M}_t)$ returns $\sum_{\ell=1}^{n_t}\mu_{t\ell}\ \eta(\mathcal{D}_i; \mathcal{T}_t,\ell)$ where $\eta(\mathcal{D}_i; \mathcal{T}_t,\ell)$ is the indicator that $\mathcal{D}_i$ is associated with leaf node $\ell$ in tree $\mathcal{T}_t$, that is, $g(\mathcal{D}_i; \mathcal{T}_t, \mathcal{M}_t) = \mu_{t\ell}$. The tree $\mathcal{T}_t$ is constructed using interior splitting rules of the form $[d_j \le C_b]$ at each branch node $b$, inducing a partition on the covariate space. The cutpoint $C_b$ of the splitting rule is assigned a uniform prior over its possible values. To avoid overfitting, each tree is given a \emph{regularization prior} which shrinks the individual leaf parameters to $0$, ensuring that each tree contributes only a small portion of the overall fit. The regularization is governed by the depth of each tree and the prior distribution of leaf parameters. The probability that each node at depth $d$ is non-terminal is given by $\gamma(1+d)^{-\beta}$ for hyperparameters $\gamma\in(0,1)$ and $\beta\in[0,\infty)$. The default prior specification \citep{Chipman2010} sets $\gamma=0.95$ and $\beta=2$, which encourages shallow trees, rarely exceeding depth $2$--$3$. The leaf parameters $\mu_{t\ell}$ are assigned iid $N(0,\sigma^2_\mu/m)$ priors, ensuring that the prior variance of $f(\bm x)$ remains constant as a function of $m$.





In extending to the multiple network setting, while considering varying $W^k$ across networks in the outcome model, we consider a semi-parametric BART model given as

\begin{align}\label{model:semi}
    Y^k_{i2} &= \sum_{t=1}^m g(\mathcal{D}^k_i; \mathcal{T}_t, \mathcal{M}_t) + (\beta_z^{k})^T \boldsymbol{Z}_{i}^k + \epsilon_{i}^k
\end{align}

where $\mathcal{D}^{k}_i = (Y_{i1}^k, G_i^k, \bX_i^k)$, $\beta_z^k \in \mathbb{R}^d$ and $\epsilon_{i}^k\sim N(0,\tau'^2)$. The coefficient $(\beta_z^{k})$ varies with network $k$ and adapts to the rotation matrix $W^k$ such that $(\beta_z^{k})^T \boldsymbol{Z}_{i}^k = (\beta_z)^T W^k \boldsymbol{Z}_{i}^k$. By using a semi-parametric BART model, we account for network-specific homophilous attributes which are not shared across the networks via the random effect $(\beta_z^{k})$. This model adopts a similar strategy to mixedBART \citep{Spanbauer2021}, where a parametric random effect term $(\beta_z^{k})^T \boldsymbol{Z}_{i}^k$ is introduced into the BART model analogous to random effect term in a linear mixed effect model. The $\sum_{t=1}^m g(\mathcal{D}^k_i; \mathcal{T}_t, \mathcal{M}_t)$ represents the population fixed effect regression relationship, shared by all of the observations, analogous to a fixed effect term in a linear mixed effect model. Therefore, this semi-parametric BART model effectively captures the nonlinear relationship between peer influence and individual behavior, which is shared by all observations, while accounting for varying rotation matrix $W^k$ across networks. For the parameter $\beta_z^k$, a multivariate normal prior distribution $\beta_z\sim N_d(0, \sigma_\beta I)$ is used. The details of our Bayesian computation and estimation algorithm are provided in  Appendix \ref{appendix:semiBART}.

\begin{algorithm}[t]\caption{LSM-BART}\label{algorithm:LSM-BART}
    \begin{algorithmic}[1]
    \For{$j = 1$ to $J$}
    \StateX \textbf{(Latent location stage)}
        \State Draw a sample of $\alpha_0^j \sim p(\cdot|\mathbf{Z}^{j-1},\beta_z^{j-1}, A)$ \text{ where } $\mathbf{Z}^{j-1} = (Z_1^{j-1},\cdots, Z_n^{j-1})$
        \State Draw a sample of $\beta_z^j \sim p(\cdot|\mathbf{Z}^{j-1},\alpha_0^{j}, A)$.
        \For{$i = 1$ to $N$}
            \State Update $Z_i^{j-1}$ given $Z_i^{j}, \alpha_0^{j},\beta_z^j$ and $A$
        \EndFor\vskip0.2cm
    \StateX \textbf{(Outcome stage) } 
        \For{$m = 1$ to $M$}
        \State Set $R^j_{m i}=y_{i2} - \sum_{q \neq m} g\left(D_i ; \mathcal{T}_q, \mathcal{M}_q\right)$ for $i=1, \ldots, n$ where $D_i = (Y_{i1}, G_i, \mathbf{Z}_i^j$).
        \State Propose a new tree $T_m^* \sim Q\left(\mathcal{T}_m \rightarrow \mathcal{T}_m^*\right)$.
        \State Compute the acceptance ratio
        \begin{align*}    
        r=\frac{\mathscr{L}\left(\mathcal{T}_m^*\right) Q\left(\mathcal{T}_m^* \rightarrow \mathcal{T}_m\right)}{\mathscr{L}\left(\mathcal{T}_m\right) Q\left(\mathcal{T} \rightarrow \mathcal{T}_m^*\right)},
        \end{align*}
        where $\mathscr{L}\left(\mathcal{T}_m\right)=p\left(\mathcal{T}_m \mid \boldsymbol{R}_m, \sigma\right) \times p\left(\mathcal{T}_m\right)$ and $\boldsymbol{R}_m=(R^j_{m1},\ldots,R^j_{mn})$.
        \State Set $\mathcal{T}_m=\mathcal{T}_m^*$ if $U \leq \min (1, r)$ where $U \sim \operatorname{Uniform}(0,1)$, otherwise retain $\mathcal{T}_m$.
        \State Sample the $\mu_{t \ell}$'s from their full conditional distributions.
        \EndFor  
        \State Update $\tau$ from its full conditional distribution.\vskip0.2cm
        \StateX \textbf{(Monte Carlo estimation stage)} 
            \For{$g \in \mathcal{G}$}
            \State Predict $\hat{Y}_{i2}(g)$ given $Z_i$, $\boldsymbol{\mathcal{T}} = (\mathcal{T}_1,\ldots,\mathcal{T}_m), $ and $\boldsymbol{\mathcal{M}} = (\mathcal{M}_1,\ldots,\mathcal{M}_m)$
        \EndFor
        \State Compute average the potential outcomes over all units as $\hat{\mu}(g) = \frac{1}{N}\sum_{i=1}^N \hat{Y}_{i2}(g)$ for all $g\in \mathcal{G}$
        \State Compute ACPI as $\tau(g,g^\prime) = \hat{\mu}(g) - \hat{\mu}(g^\prime)$ for all $g, g^\prime\in \mathcal{G}$
    \EndFor
    \end{algorithmic}
\end{algorithm}

\normalcolor

\subsection{Two-step Estimation}

We estimate our model using a Markov chain Monte Carlo (MCMC) algorithm consisting of two steps: 1) the update of latent locations and related LSM parameters, 2) the update of the parameters in the outcome model. Let $\btheta_z$ and $\btheta_y$ be the vectors
of parameters of network (LSM) and the outcome model, respectively. The full conditional distributions for each parameter are
\begin{align}\label{eq:posterior}
&p\left(\btheta_y \mid \boldsymbol{Y}_{2}, \boldsymbol{Y}_1, \mathbf{X},\btheta_z,A\right) \propto p\left(\boldsymbol{Y}_{1}, \boldsymbol{Y}_{2}, \mathbf{Z}, \mathbf{X}, A \mid \mathbf{Z}, \btheta_z, \btheta_y\right) p\left(\btheta_y\right)\nonumber\\
&p\left(\mathbf{Z}, \btheta_z \mid \boldsymbol{Y}_{2}, \boldsymbol{Y}_1, \mathbf{X},\btheta_y, A\right) \propto p\left(\boldsymbol{Y}_{1}, \boldsymbol{Y}_{2}, \mathbf{Z}, \mathbf{X},A \mid \mathbf{Z}, \btheta_z, \btheta_y\right) p\left(\mathbf{Z}, \btheta_z\right)
\end{align}
respectively. Within the Bayesian framework, the joint posterior distribution of $\mathbf{Z}$ and $\btheta_z$ is in part informed by the update in outcome model. This may raise concerns when 1) connections between individuals were established before the behaviors are disseminated, and 2) model misspecification in the relationship between outcome and latent locations results in incorrect causal estimates. This is similar to model feedback \citep{HOSHINO2008, Zigler2013} where the posterior distribution of the parameters in the propensity score model is partially informed by the outcome model.

 
To address these concerns, we replace the full conditional distribution $\eqref{eq:posterior}$ with an approximate conditional distribution given as
\begin{align*}
p\left(\mathbf{Z},\btheta_z \mid \boldsymbol{Y}_{2}, \boldsymbol{Y}_1, \mathbf{X},\btheta_y, A \right) \propto  p\left(A \mid \mathbf{Z},  \btheta_z \right) p\left(\mathbf{Z},\btheta_z\right) 
\end{align*}
employing a similar strategy as in \cite{HOSHINO2008,kaplan2012,mccandless2010,Bayarri2009}. 

Samples are drawn from the posterior predictive distribution of the potential outcomes $Y_{i2}(g)$ for each $i$. Subsequently, for each sample, we compute the average potential outcome $\mu(g)$ by averaging $Y_{i2}(g)$ across all nodes. The ACPI is then computed by comparing the average potential outcomes at different levels. A description of the estimation process using the LSM is provided in Algorithm \ref{algorithm:LSM-BART}, and the  algorithm with the HLSM is available in the Appendix.

\section{Simulation Study}\label{sec:simulation}


To assess the performance of our proposed framework for estimating the ACPI, we conduct a series of three simulation studies where we compared our method to a naive estimator that does not adjust for latent homophily and an oracle estimator assuming that the latent homophilous confounder $U_i$ is observed. In the first simulation setting in Section \ref{sec:sim1}, we assess how the strength of network homophilly, as measured by the association between the latent attribute and the likelihood of tie formation, impacts ACPI estimation. In the second set of simulations in Section \ref{sec:sim2}, we investigate how the strength of unmeasured confounding, quantified by the dependence between the latent attribute and the baseline behavior, may impact the estimation. The first two simulation settings focus on single network data. In the last simulation setting  in Section \ref{sec:sim3}, we investigate ACPI estimation with multiple networks, similar to our motivating example. 

We first present our data generating process for a single network using the LSM. To incorporate homophily through clusters, we generate attributes $\bU_i\in\mathbb{R}^2$ for each node $i$ using Gaussian mixture model with $J$ components, such that the $j$th component has a mean of $\mu_j$ and a diagonal variance-covariance matrix of $\sigma^2_jI$, as $\boldsymbol{U}_i \sim \sum_{j=1}^J \phi_j MVN_2(\mu_j, \sigma^2_jI),$
where $\phi_j$ is the mixture component weight subject to the constraint $\sum_{j=1}^J \phi_j=1$. In all simulations, we allocate each node to a component by sampling the corresponding component $j$ using a $Multinomial(n; p_1, \cdots, p_J)$, where $p_j = 1/J$. We fix $J=4$ and the mean vector for each component of the Gaussian mixture model as $\mu_1=(-1.5,-1.5)$, $\mu_2=(1.5,-1.5)$, $\mu_3=(-1.5,1.5)$, and $\mu_4=(1.5,1.5)$. We set $\phi_j$ to $1/J$ to assign equal weight to each component. Next we generate an undirected network following LSM as specified in \eqref{eq:lsm} assuming $U_i=Z_i$ with $\alpha_0 = 6$ and $\beta_z = 5$. In all simulations, we fix $\beta_z$ as we demonstrate variations in homophily strength by varying $\sigma_j$ instead. Further, for all simulations with a single network we fix the number of nodes $n$ to 100. 
To generate multiple networks for simulations in Section \ref{sec:sim3}, we use the HLSM in \eqref{eq:HLSM} by setting $\alpha_0^k = 3$ in \eqref{eq:HLSM}. Each network is generated following the same strategy as in the single network setting using a Gaussian mixture model with $n_k = 40$ and $J=4$ where the number of nodes within each network was chosen to mimic our motivating example. 




Given our 100-node networks, we generate individual behaviors at baseline as 
 $Y_{i1} = 5 + \beta_{u_1}U_{i1} + \beta_{u_2}U_{i2} + \epsilon_{i1}$,
where $\epsilon_i \sim N(0,1)$.  We further specify a nonlinear outcome model to generate individual behaviors at follow-up as
$Y_{i2} = 10 + 3Y_{i1} + 10 \text{sigmoid}(10 (G_i - 5)) + \beta'_{u_1} U_{i1} + \beta'_{u_2} U_{i2} + \epsilon_{i2}$,
where $\text{sigmoid}(x) = 1/(1+ e^{-x})$. 
 With these true underlying values, the population average potential outcome  is $\E(Y_{i2}(g)) = 25 +10 \text{sigmoid}(10 (g - 5))$.  Details of the data generating process in the multiple networks setting are provided in Appendix \ref{appendix:HLSM}

To investigate the influence of homophily on ACPI, we vary $\sigma_j$ in the Gaussian mixture model for $U_i$ in Section \ref{sec:sim1}, whereas we fix $\sigma_j = 0.3$ in the other two simulation settings. Similarly, to assess the impact of varying strength of association of $U_i$ with $Y_{i1}$ and $Y_{i2}$ 
(simulations in Section \ref{sec:sim2}), we vary $\beta_{u_1}$, $ \beta_{u_2}$, $\beta'_{u_1}$ and $\beta'_{u_2}$. For other simulations they are fixed at $\beta_\mathbf{u}=(\beta_{u_1},\beta_{u_2})=(0.5, 0.5)$ and $\beta'_\mathbf{u}=(\beta'_{u_1},\beta'_{u_2})=(10,10)$. Finally, we vary the number of networks in Section \ref{sec:sim3}.

To sample from the posterior distribution of the latent locations $Z_i$ using the LSM, we set the hyperparameters as $\sigma_z = 5, \ \mu_\alpha=0, \ \mu_\beta=0,\  \sigma_\alpha=10, \ \sigma_\beta=10$. For the multiple networks setting, we set $\sigma_{z}^\prime = 5, \ \mu_{\alpha^\prime} = 0, \ \mu_{\beta^\prime} = 0, \ \sigma_{\alpha^\prime}=10, \ \sigma_{\beta^\prime}=10$. We next estimate the posterior mean (with 95\% posterior intervals) of the ACPI using either BART (for single network) or Semi-BART (for multiple networks) with 10000 MCMC iterations with 5000 burn-in. We set the number of trees $m$ to 200. While it is possible to include the additional confounders $X_i$ that are independent of network tie formation yet associated with $Y_{i1}$ and $Y_{i2}$, our simulation excludes $X_i$ to precisely investigate how unobserved homophily can bias ACPI estimation and whether these latent homophilous attributes can be estimated by conditioning on the observed network data. Further details on the estimation can be found in Appendix \ref{appendix:HLSM}. We use the \emph{HLSM} \textit{R} package to estimate latent locations.

The average potential outcome $\mu(g)$ is calculated by averaging $Y_{i2}(g)$ across all nodes, as outlined in Algorithm \ref{algorithm:LSM-BART}. This computed $\mu(g)$, is also examined in the simulation studies. It is important to note that while $\mu(g)$ can be computed for all $g \in \mathcal{G}$, the primary estimand of interest, ACPI, involves comparisons between the average potential outcomes $\mu(g)$ of two distinct treatment levels, $g$ and $g'$. To evaluate the ACPI estimation with two levels, we partition the range of interest into 10 equally spaced intervals, represented as $(g_1, \ldots, g_{10})$. We then compute the ACPI, defined as $\tau(g_{q+1}, g_q) = \mu(g_{q+1}) - \mu(g_q)$, for each consecutive pair $q=1,\ldots,9$. We set this range for peer behaviors from 4 to 6, such as $(g_1, \ldots, g_{10}) = (4, 4.22, \ldots, 5.77, 6)$, which covers the 95\% range of observed peer behaviors in all simulation settings.

For each data generating model, we replicate the simulation 10 times, and three different criteria for evaluation are computed at each replication. The first criteria is the root mean squared error (RMSE) in estimating the ACPI with the posterior mean. The second criteria is the coverage or the proportion of 95\% credible intervals that capture the true value of the ACPI. The third criteria is the average length of the 95\% credible interval for the ACPI. Then, our evaluation criteria are averaged over 10 replications and the entire grid. Specifically, $RMSE = \sqrt{\sum_r \sum_q \left( \hat{\tau}^r(g_{q+1}, g_{q}) - \tau^r(g_{q+1}, g_{q})\right)^2/RQ}$, where $R$ is the number of replications and $Q$ is the number of grid points in the range $(Q=10)$. The $\hat{\tau}^r(\cdot)$ represents the posterior mean at the $r$th replicate. The coverage and length are defined in a manner similar to RMSE.

 \begin{figure}[!t]
    \centering
    \includegraphics[width=1 \linewidth]{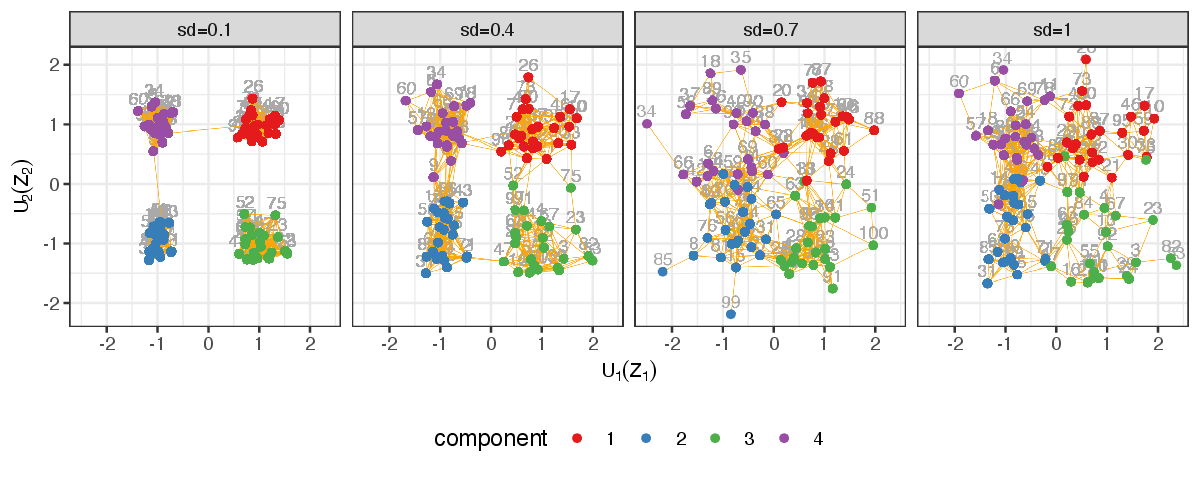}
    \caption{Simulation 1: The generated networks from data generating process with $\sigma_j \in (0.1, 0.4,0.7,1)$.
}
    \label{fig:sim_sd}
\end{figure}

\subsection{The Strength of Homophily and ACPI}\label{sec:sim1}

\begin{figure}[t]
    \centering
    \includegraphics[width=1 \linewidth]{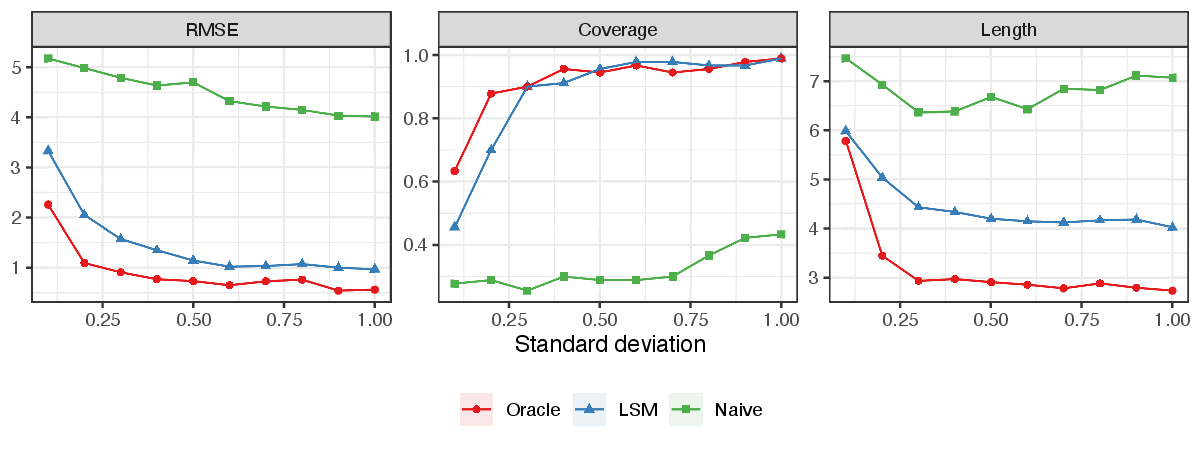}
    \caption{Simulation 1: The RMSE, coverage, and length of ACPI estimation from oracle, LSM, and naive estimator are evaluated based on 10 replications of the simulation at each standard deviation value.}\label{fig:sim_sd_rmse}
\end{figure}

\begin{figure}[!t]
    \centering
    \includegraphics[width=1 \linewidth]{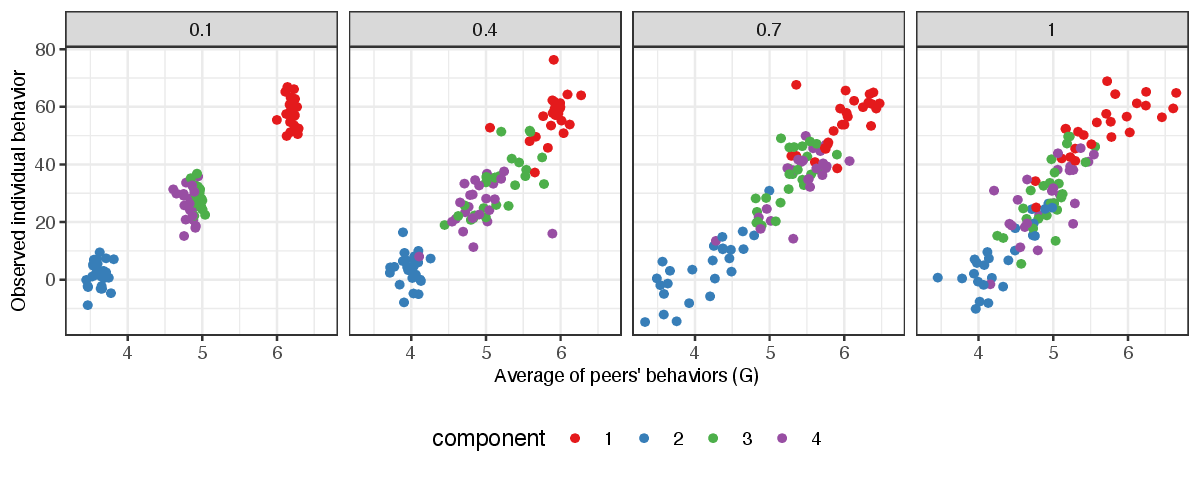}
    \caption{Simulation 1: The scatter plot of observed individual behavior at follow-up ($y_{i2}$) versus the (observed) average of peers' behaviors with $\sigma_j \in (0.1, 0.4,0.7,1)$.}\label{fig:sim_sd_net}
\end{figure}

In this simulation study, we examine the impact of the homophily's strength on ACPI estimation, with the hypothesis that strong clustering should induce strong homophily as well as increased oppotunities for influence \citep{Warren2020,Malik2016}. When we generate latent locations, the variance of latent locations within the same component is controlled by $\sigma_j$, which can be interpreted as the strength of the homophily. We evaluate the impact of these different homophily strengths within a component by varying $\sigma_j \in (0.1, 0.2, \ldots, 0.9, 1)$.  Examples of generated networks with $\sigma_j = (0.1, 0.4, 0.7, 1)$ are illustrated in Figure \ref{fig:sim_sd_net}. When $\sigma_j=0.1$, nodes within the same component are more likely to be connected while connections between components are limited. Additionally, smaller variations or closer proximity between nodes within a component imply that the nodes share more similar attributes. In contrast, at $\sigma_j=1$, nodes still tend to connect within their component, maintaining a homophilous network, but the distinction between components becomes less apparent, leading to increased connections between components.

The average RMSE, coverage, and interval length of the ACPI estimation performance is represented in Figure \ref{fig:sim_sd_rmse}. Both the LSM and Oracle exhibit similar performances in terms of RMSE and coverage, with the LSM achieving significantly narrower interval lengths compared to the Naive estimator. This suggests that, regardless of the strength of homophily, including latent locations leads to an improvement in mitigating bias in the estimation of ACPI while also narrowing the ACPI credible intervals. Another observation is that both Naive and LSM estimators perform poorly when standard deviations are small ($\sigma_j \leq 0.3$), and the performance is stabilized for ($\sigma_j > 0.3$) across all criteria.

 \begin{figure}[t]
    \centering
    \includegraphics[width=1 \linewidth]{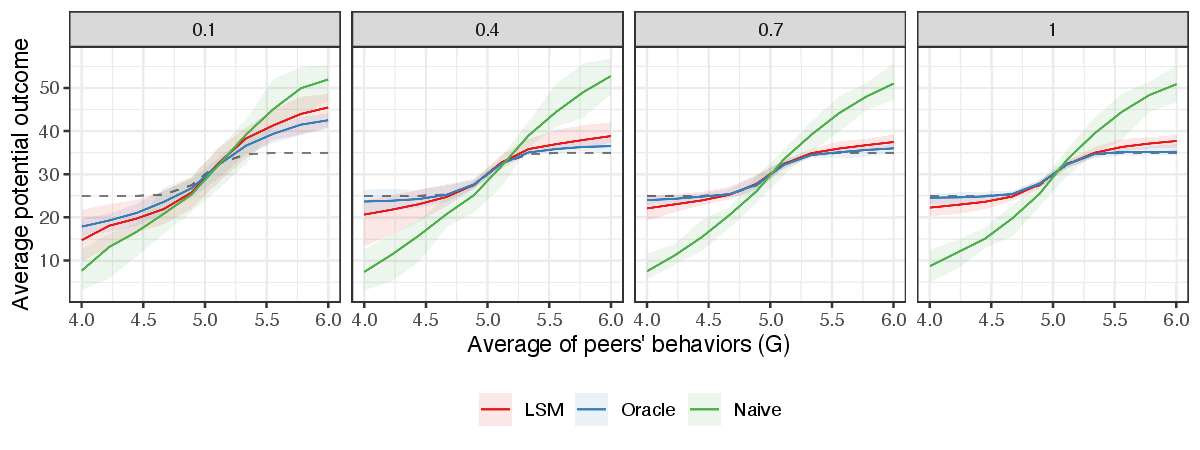}
    \caption{Simulation 1: The estimated average potential outcome  $\mu(g)$ from oracle, LSM, and naive estimator with standard deviations $\sigma_j$ of 0.1, 0.4, 0.7, and 1.0. The 95\% confidence interval is also represented by shading along with the true average potential outcome  (dotted line) obtained from 10 replications.
}\label{fig:sim_sd_apo}
\end{figure}

The underperformance with smaller standard deviations is primarily attributed not to estimation inaccuracies but to a violation of the positivity assumption. To see this violation, Figure \ref{fig:sim_sd_net} presents the observed individual behaviors colored by each component, and Figure \ref{fig:sim_sd_apo} shows the estimated average potential outcomes with $\sigma_j$ = (0.1, 0.4, 0.7, 1). Given $\sigma_j=0.1$, the observed behaviors in component 1 are less likely to be similar to those in component 2 in Figure \ref{fig:sim_sd_net}. That is, in the presence of strong homophily, behaviors are more distinct across different components in Gaussian mixture model as baseline behaviors are determined by the homophilous attributes $\mathbf{U}_i$. Also, there exist ranges in which peers' behaviors are not observed leading to biases in average potential outcome estimation. It is important to note that behaviors observed from each component do not cover the entire range of $g$. However, in Figure \ref{fig:sim_sd_apo}, our oracle and LSM estimators effectively recover true average potential outcome, except for strong homophily such as $\sigma_j=0.1$. Conversely, the estimated average potential outcome  from Naive estimator remains constant irrespective of the strength of homophily, exhibiting the largest bias. This consistent performance arises due to the Naive estimator's ignorance of the homophily effect.

\subsection{Coefficient of Unmeasured Confounding and ACPI}\label{sec:sim2}

\begin{figure}[t]
    \centering
    \includegraphics[width=1 \linewidth]{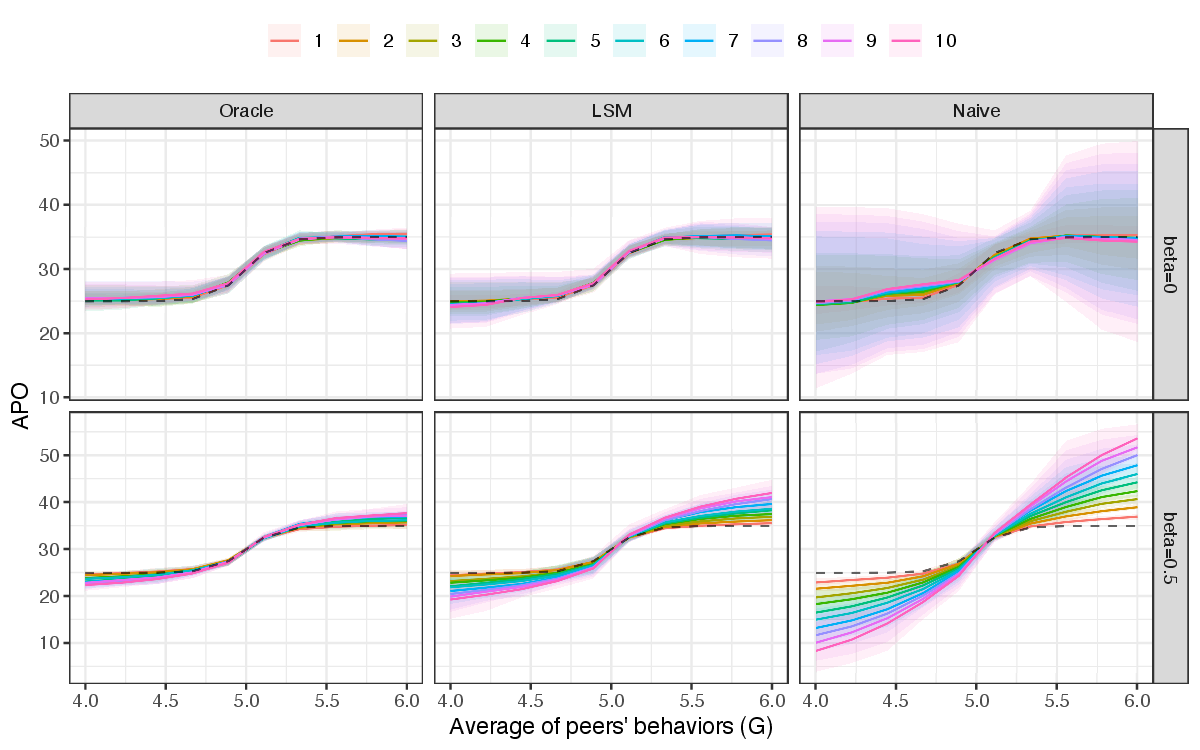}
    \caption{Simulation 2: The estimated average potential outcome from oracle, LSM, and naive estimator varying $\beta_\mathbf{u}^\prime\in(1,\ldots,10)$, represented by different colors. The population average potential outcome $\mu(g)$ is outlined with a dotted line. Results for $\beta_\mathbf{u} = 0$ are shown in the first row, and for $\beta_\mathbf{u} = 0.5$ in the second row. The 95\% confidence interval is also represented by shading along with the true average potential outcome (dotted line) obtained from 10 replications.}
    \label{fig:coef_APO}
\end{figure}

\begin{figure}[!t]
    \centering
    \includegraphics[width=1 \linewidth]{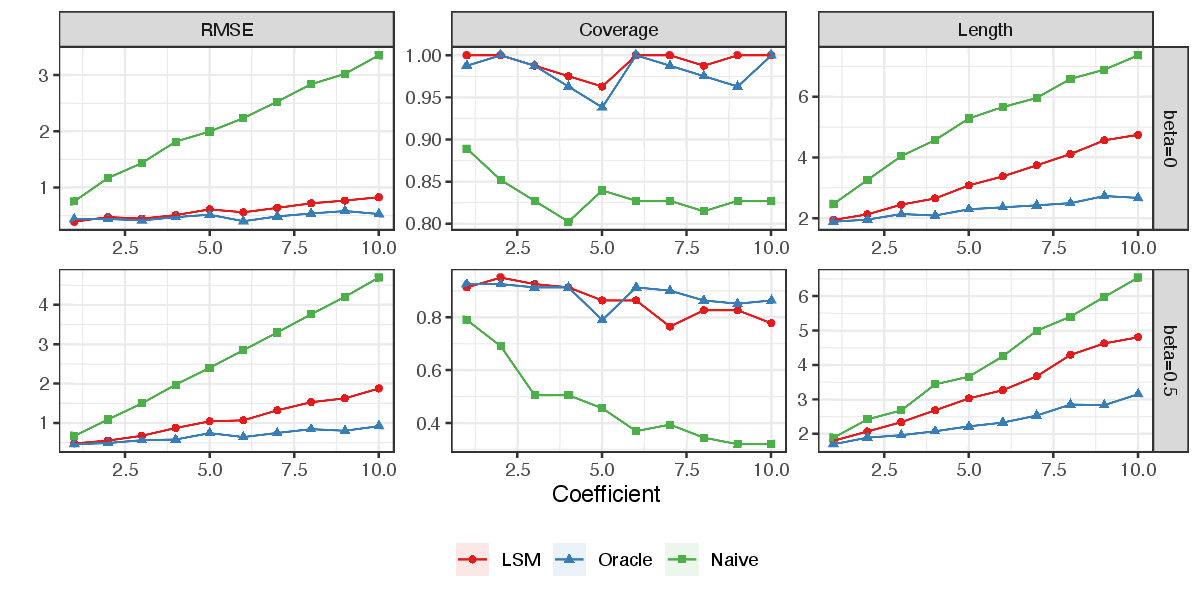}
    \caption{Simulation 2: The RMSE, coverage, and length of ACPI estimation from oracle, LSM, and naive estimator are evaluated based on 10 replications of the simulation at each coefficient value $\beta^\prime_{\mathbf{u}}$. Results are shown with $\beta_\mathbf{u} = 0$  (top) and $\beta_\mathbf{u} = 0.5$ (bottom).}
    \label{fig:coef_RMSE}
\end{figure}

In our second simulation study, we explore impact of the effect size of latent homophilous attributes in behaviors at baseline and follow-up on the estimation of ACPI by varying coefficients of $U_i$ in outcome models for $Y_{i1}$ and $Y_{i2}$. We consider two scenarios: i) $U_i$ has no influence on $Y_{i1}$ with $\beta_{\mathbf{u}} = (\beta_{u_1}, \beta_{u_2})=(0,0)$ and $\epsilon_{i2} \sim N(0,2^2)$, and ii) $U_i$ has influence on $Y_{i1}$ with $\beta_{\mathbf{u}}= (\beta_{u_1}, \beta_{u_2}) = (0.5,0.5)$ and $\epsilon_{i1} \sim N(0,1)$. The first scenario indicates that the latent homophilous attributes $\mathbf{U}_i$ are independent of $Y_{i1}$ implying no arrow from from $U_i$ to $Y_{i1}$ in Figure \ref{fig:DAG} while keeping an arrow from  $U_i$ to $Y_{i2}$. This implies that behaviors at baseline are randomly assigned, but the network ties still depend on latent homophilous attributes. The standard deviation of $\epsilon_{i2}$ in the first scenario is increased to ensure that the range of $Y_{i1}$ and subsequently $G_i$ is similar to their range in the second scenario. For both scenarios, we also vary the value of $\beta'_{u_1}=\beta'_{u_2}\in(1,2,\cdots,9,10)$, to investigate the influence of the effect size of the latent homophilous attribute in the behaviors at baseline on the estimation of average potential outcome as well as ACPI.

The results of average potential outcome estimation are presented in Figure \ref{fig:coef_APO} for the first scenario (top) and the second scenario (bottom). In the first scenario where $Y_{i1}$ are randomly assigned, the magnitude of the coefficient $\beta'_u$ in the follow-up outcome model does not have any effect on the average potential outcome estimation across all three methods. Also, the estimated average potential outcomes from all methods capture ground truth well. However, the confidence interval obtained from 10 replications in the Naive estimator, is significantly wider, especially as $\beta'_u$ increases. This observation suggests that ignoring the latent homophilous attribute results in a wider confidence interval when $\bU_i \Perp Y_{i1}$, thereby indicating increased uncertainty in the estimation. In this result, ignoring latent attributes does not contribute to significant bias in the estimation of the average potential outcome when peers' behaviors at follow-up are assigned randomly; instead, it merely increases the variance. On the other hand, in the second scenario (bottom panel in Figure \ref{fig:coef_APO}), where the latent homophilous attributes $\bU_i$ is involved in the behavior at baseline, Naive estimator shows that the the bias in estimating average potential outcome  increase as value of the $\beta'_\mathbf{u}$ increases. However, our proposed LSM estimator is effective in reducing this type of bias by adjusting for the latent locations. Although the bias in the estimated average potential outcome slightly increases as $\beta'_\mathbf{u}$ increases, it remains markedly lower compared to the Naive estimator. Moreover, the performance of LSM estimator is comparable to the result of Oracle estimator. This result demonstrates that adjusting for the latent locations $Z_i$ effectively mitigates bias caused by homophily, especially when the baseline behavior is also impacted by latent homophilous attributes.

Next, we compute the average RMSE, coverage, and interval length, represented for ACPI estimation in Figure \ref{fig:coef_RMSE} for the first scenario (top) and second scenario (bottom). For both scenarios, the RMSE from the Naive estimator strictly increases as $\beta'_{\mathbf{u}}$ increases, while the average coverage is consistently lower than that of its competitors. Additionally, the CI length from Naive estimator is consistently larger than those of both the LSM and the Oracle method,regardless of the values of $\beta'_{\mathbf{u}}$. Especially, RMSE from LSM estimator is significantly reduced compared to that from Naive estimator.

\subsection{Number of Networks and ACPI: Extension to HLSM}\label{sec:sim3}

In this section, we extend our approach to the multiple networks setting and investigate the impact of the number of networks on the estimation of ACPI. We varied the number of networks  $k \in (2, 4, \ldots, 12, 14)$ and generated network data from the HLSM given in \eqref{eq:HLSM}. 

\begin{figure}[!t]
    \centering
    \includegraphics[width=1 \linewidth]{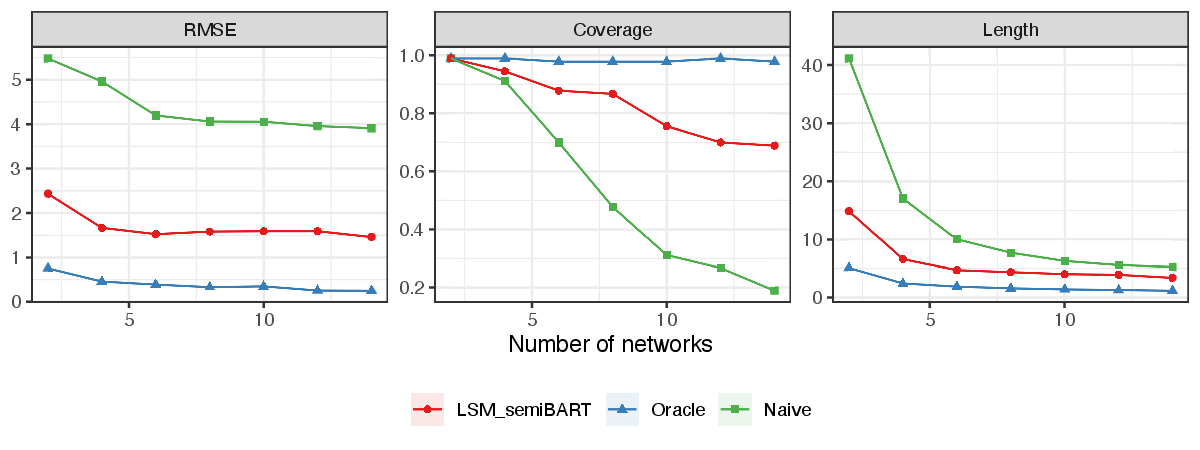}
    \caption{Simulation 3: The RMSE, coverage and length of ACPI are evaluated based on 10 replications of the simulation varying the number of networks with $k\in(2,4,\ldots,12,14)$.}    
    \label{fig:HLSM_RMSE}
\end{figure}

We compute the average RMSE, coverage, and interval length for each $k$, represented in Figure \ref{fig:HLSM_RMSE}. For all evaluated criteria, the Naive estimator demonstrates inferior performance, exhibiting the highest RMSE and lowest coverage. In contrast, the LSM-semiBART estimator significantly reduces the RMSE and provides improved coverage. This indicates that incorporating adjustments for latent locations effectively reduces bias associated with homophily and decreases uncertainty within a multi-network context. Another finding is that for all estimators, both RMSE and length remain relatively consistent when $K\geq 6$. This indicates that a sample comprising 6 networks ($N = n_k \times K = 40 \times 6 = 240$) is adequately large to ensure robust estimation of the ACPI in our simulation setting. 

Unlike the simulations in single network settings described in Sections \ref{sec:sim1} and \ref{sec:sim2}, the coverage of the LSM-semiBART estimator is smaller compared to that of the Oracle estimator. This decrease may be attributed to the expansion of the parameter space. For a fixed set of nodes, as the number of networks ($k$) increases, the dimension of the parameter $\beta^k_\mathbf{z} = (\beta^k_{z_1}, \beta^k_{z_2})$ also increases, accounting for variations in transformations between networks. This expanded parameter space likely results in a decrease in coverage as $k$ increases. Nevertheless, the LSM-semiBART estimator significantly outperforms the Naive estimator across all evaluation criteria, and the RMSE remains overall consistent, regardless of the number of networks $k$. This performance indicates that our proposed LSM-semiBART model effectively estimates the shared ACPI across networks by accounting for the transformed latent locations across different networks.




\section{Real Data Analysis}\label{sec:real}
To investigate the causal peer influence effect on beliefs about mathematics using advice-seeking networks, we apply our proposed method to estimate the ACPI. Recall from Section \ref{subsec:mot}, our data include 14 advice-seeking networks among elementary school staff in a school district. We hypothesize that teachers' beliefs in 2013 were shaped not only by their own beliefs but also by those of their peers in 2012. For this analysis, from a total of 14 schools, we selected 11 schools with more than 20 respondents (nodes) and assumed an undirected relationship, where all ties are considered to be reciprocated. We use the advice-seeking network data collected in Spring 2013, as teachers were asked to nominate others based on their interactions over the past year.

Following the LSM-semiBART approach for multiple networks to assess the average potential outcome $\mu(g)$, we define the range for $g$ to be from $15.6$ to $22.6$, which empirically encompasses the 95\% range of observed average peer belief scores. Then we partition this range into 10 equally spaced intervals such as $(g_1, g_2, \cdots,g_ 9, g_{10}) = (15.6, 16.0, \cdots, 22.2, 22.6)$ and evaluate the average potential outcome $\mu(g_q)$ for $q=1,\ldots,10$ and ACPI $\tau(g_{q+1}, g_{q}) = \mu(g_{q+1}) - \mu(g_{q})$ for $q=1,\ldots,9$. 

\begin{figure}[t]
    \centering
    \includegraphics[width=1 \linewidth]{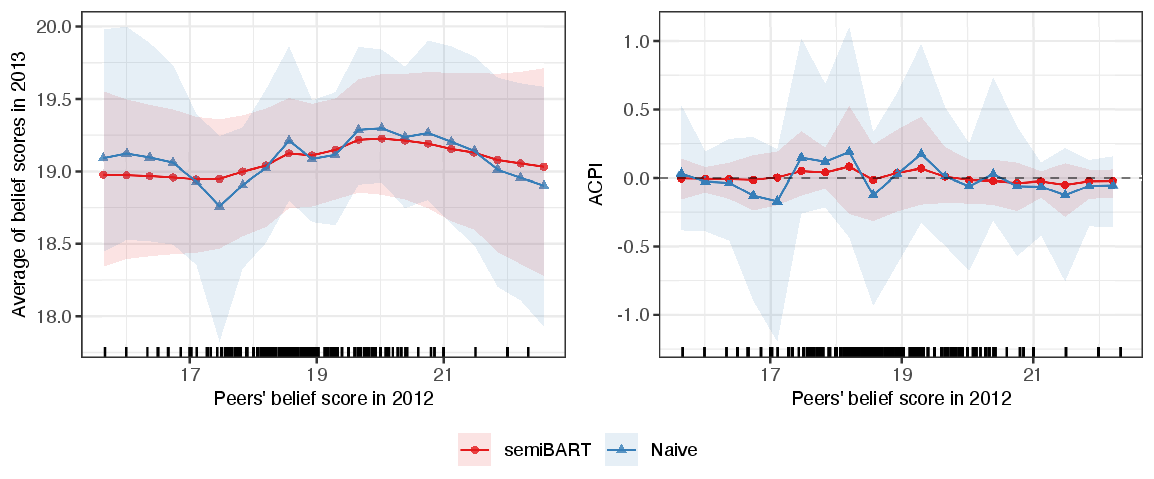}
    \caption{The estimated average belief scores in 2013 across all school from the Naive and LSM-semiBART estimator is represented with 95\% credible interval (left). The estimated ACPI is depicted with 95\% credible interval from each estimator (right). The rug plot shows the observed distribution of peers' belief score.}
    \label{fig:real}
\end{figure}

The left panel of Figure \ref{fig:real} illustrates the estimated average belief scores (simply referred to as average belief scores) in 2013 versus the average peers' belief score $g$ in 2012, estimated from the Naive estimator and LSM-semiBART. These average belief scores are calculated across all schools, with detailed estimates for each school provided in Figure \ref{fig:real_all} in the appendix. 
The Naive estimator exhibits a fluctuating pattern in the average belief scores, along with a wider 95\% credible interval, compared to the LSM-semiBART estimator. This pattern implies that neglecting the homophily effect introduces bias  and increases uncertainty in the estimating average belief scores.

The right panel of Figure \ref{fig:real} represents the estimated ACPI $\tau(g_{q+1},g_{q})$ for $q=1,\cdots,9$ from Naive and LSM-semiBART estimators. The Naive estimator leads to ACPI being either overestimated or underestimated, with fluctuations. Additionally, neglecting the latent homophily results in wider credible intervals, thus diminishing the reliability of the estimation. Although the impact of latent homophily on estimating the ACPI in our dataset is not markedly significant, it helps mitigate the bias introduced by overlooking the latent homophily and enhances the precision of our ACPI estimates.

Finally to ensure that the HLSM specification provides a reasonable approximation of the latent homophilous attribute, we perform goodness of fit checks using the posterior predictive distributions of the networks. The visualization of the network summaries (tie density and transitivity) of the predicted networks in Figure \ref{fig:HLSM_fit_density} demonstrates that the network ties from the prediced networks closely resemble those of the observed networks.


In our analysis of real data, we observed a lack of significant ACPI. Previous studies, which employed different models and utilized latent locations in various ways with directed networks \cite{Spillane2018, Sweet2020}, also reported minimal evidence of a peer influence effect on student-centered beliefs on average across schools. This insignificance of ACPI may be attributed to the limited duration of the observation period. 


\section{Discussion}\label{sec:discussion} 

Social network structure is caused by either observed or unobserved homophily, which can result in biased estimates of causal peer influence. Our proposed LSM-BART (LSM-semiBART) estimator demonstrated that this bias can be mitigated by adjusting for the latent locations of each node. We employed BART to estimate the nonlinear relationship between individual behavior and peers' behaviors and showed how our framework can be extended from a single network to a multiple networks setting. Our estimation process leverages a two-step Bayesian procedure, which enables the quantification of uncertainty in both latent location and potential outcome estimation. Through a series of simulation studies, we found that neglecting latent homophily led to biased estimates of the ACPI, whereas adjusting for latent locations effectively mitigated this bias. In the context of multiple networks, the semi-parametric BART model effectively explained the variation in transformations across networks as well as the nonlinear relationship between individual behavior and peers' behaviors.

In our real data analysis, we estimated the ACPI to gain insights into changes in teachers' beliefs about mathematics that resulted from peer influence. Although the impact of latent homophily on the ACPI was not significant, we observed that overlooking latent homophily can lead to either underestimation or overestimation of the ACPI, accompanied by considerable uncertainty. Given the potential benefits of accurately identifying the ACPI for enhancing teaching quality, these findings highlight the importance of incorporating latent locations to better understand homophily among teachers and applying appropriate regression models. While node-level covariates could potentially account for part of the latent homophily, we focus more on estimating latent locations using solely the observed network to account for latent homophily.

Our framework focuses on identifying the effects of peer behaviors on individual behaviors, where both are measured on a continuous scale. This approach represents a generalized version of the study on causal spillover effects in social networks \citep{Forastiere2021,Forastiere2022JMLR} by transitioning from a binary to a continuous treatment covariate. Thus, our framework can be applied to spillover effect studies, especially when latent homophily exists and the relationship between the outcome and spillover effect is nonlinear. For future work, our framework can be extended to identify heterogeneous causal peer influence effects using BART. While our current focus is on the nonlinearity between individual behaviors and their peers' behaviors, the complex interactions among peer influence, confounders, covariates, and treatment can be identified by applying BART.



\section{Acknowledgement}
We thank Jim Spillane and the Distributed Leadership Studies at Northwestern University \url{http://www.distributedleadership.org}, including the NebraskaMATH study, for the data used in this paper.

\newpage

\bibliographystyle{unsrtnat}
\bibliography{references} 

\appendix

\begin{figure}[!t]
    \centering
    \includegraphics[width=1 \linewidth]{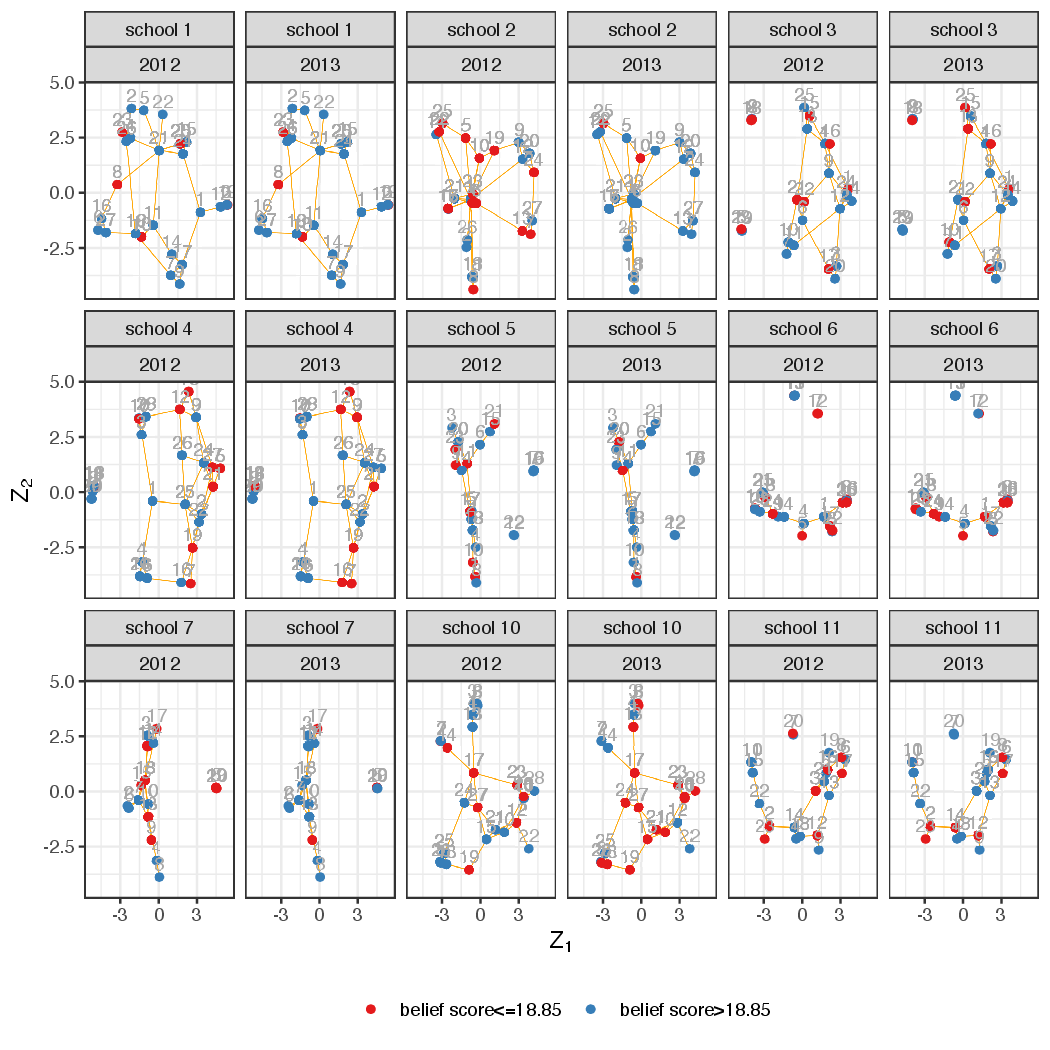}
    \caption{The teachers' scores about student-centered beliefs for 9 schools with more than 20 teachers in both 2012 and 2013. The belief scores are represented visually by coloring each node based on high/low belief scores, using a cut-off of $17$ for effective visualization.}
\label{fig:net_real_obs_all}
\end{figure}

\section{The detail of the real data}
\begin{table}[!h]
  \centering
  \begin{tabular}{ll}
    \toprule
    \textbf{Item} & \textbf{Statment}  \\
    \midrule
    1 & Teachers should encourage students to find their own solutions to math problems even if they are inefficient\\[1ex] 
    2 & Teachers should allow students to figure out their own ways to solve simple word problems \\[1ex] 
    3 & The goals of instruction in mathematics are best achieved when students find their own methods for solving problems  \\[1ex] 
    4 & Most students can figure out ways to solve many mathematics problems without any adult help  \\[1ex] 
    5 & Mathematics should be presented to children in such a way that they can discover relationships for themselves \\
    \bottomrule
  \end{tabular}
  \caption{Teachers in Auburn Park were surveyed regarding their beliefs about teaching mathematics. They were asked to indicate the extent to which they agreed with the statements above.}
  \label{tab:survey}
\end{table}

\newpage

\section{Proof for identification}

Theorem \ref{thm:identification} indicates that the average potential outcome $Y_{i2}(g)$ can be obtained by an unbiased estimator of conditional outcome mean with proxy variable $Z_i$. This identification enables us to estimate the causal peer influence, which is determined by the function of potential outcome. Under assumption \ref{assump:consistency}-\ref{assump:ignorability},





\begin{align*}
\E \left[Y_{i2}(g)|\mathcal{A}_n\right] &= \E\left[\E \left[Y_{i2}(g) | Y_{i1} = y \bX_i = \bx, U_i = u \right]|\mathcal{A}_n\right]\\
&= \E\left[ \E \left[Y_{i2}(g) | Y_{i1} = y, G_i = g, \bX_i = \bx, U_i = u \right]|\mathcal{A}_n \right]\quad \text{by Assumption \ref{assump:ignorability}} \\
&= \E\left[\E \left[\E \left[Y_{i2}(g) | Y_{i1} = y, G_i = g, \bX_i = \bx, U_i = u \right] | Z_i = z\right]|\mathcal{A}_n\right] \\
&= \E\left[\E \left[Y_{i2} (g)| Y_{i1} = y, G_i = g, \bX_i = \bx, Z_i = z \right] |\mathcal{A}_n\right] \quad \text{by tower property; $U_i$ is $Z_i$-measurable} \\
&= \E \left[\E \left[Y_{i2} | Y_{i1} = y, G_i = g, \bX_i = \bx, Z_i = z \right]|\mathcal{A}_n \right]  \quad \text{by Assumption \ref{assump:consistency} and Assumption \ref{assump:Nei_interference}}\\
&= \E \left[Y_{i2} | Y_{i1} = y, G_i = g, \bX_i = \bx, Z_i = z \right]  \quad \text{by Assumption \ref{assump:proxy}}\\
&= \E \left[Y_{i2} | Y_{i1} = y, G_i = g, \bX_i = \bx, Z^*_i = z^* \right]  \quad \text{by $Z_i^* = WZ_i$ which is an invertible transformation}\\
\end{align*}

Given that $U$ is $Z$ measurable, tower property is $E[X|Z] = E[E[X|Z]|U] = E[X|U]$. This proof adopts the proof presented by \cite{Cristali2022}, which also relies on the tower property.

\section{Prior specification and posterior computation in the semi-parametric BART}\label{appendix:semiBART}

For the tree structures, we use the default prior for the BART model. Our Gibbs sampling algorithm uses a Metropolis-within-Gibbs strategy to iteratively update ($\mathcal{T}_i, \mathcal{M}_t$) for $t=1,\ldots,m$. Sampling from ($\mathcal{T}_i; \mathcal{M}_t$) is equivalent to sampling from
$$
p\left(\mathcal{T}_t, \mathcal{M}_t \mid \boldsymbol{R}_t, \tau\right)
$$
where $\boldsymbol{R}_t=\left(R_{t 1}, \ldots, R_{t n}\right)^{\top}$ with $R_{t i} = Y_i - \sum_{j \neq t} g\left(D_i^k ; \mathcal{T}_j, \mathcal{M}_j\right) - (\beta_z^{k})^T \boldsymbol{Z}_{i}^k$. This allows us to use the existing Bayesian backfitting algorithm \citep{Hastie2000,Chipman2010} to update the $\left(\mathcal{T}_t, \mathcal{M}_t\right)$. 

We use multivariate Normal prior distribution $N_d(0, \sigma_\beta  I)$ for $\pmb\beta^k_z$. The conditional posterior for the $\pmb\beta_z^k$ is
$$
\pmb\beta_z^k \sim N_d \left(\Psi^k \sum_{i=1}^{n_k} \boldsymbol{Z}_{i }^k \left(Y_{i2}^k - \mu_{i}^k\right)/\tau^{\prime2}, \Psi^k\right)
$$

where $\Psi^k=\left( \sum_{i=1}^{n_k} \boldsymbol{Z}_{i}^k \boldsymbol{Z}_{i}^{kT}/\tau^{\prime 2} + (\sigma_\beta I) ^{-1} \right)^{-1}$ and $\mu_{i}^k =  Y_i - \sum_{j \neq t} g\left(D_i^k ; \mathcal{T}_j, \mathcal{M}_j\right) $.

\begin{algorithm}[t]\caption{A single iteration for HLSM-BART}\label{algorithm:HLSM-BART}
    \begin{algorithmic}[1]
    \StateX \textbf{(Latent location stage)}
    \For{$k = 1$ to $K$}
        \State Draw a sample of HLSM parameters $\theta_z^k \sim p(\cdot|\mathbf{Z}^k, A^k)$ \text{ where } $\mathbf{Z}^k =(Z_1^k,\cdots, Z_n^k)$
        \For{$i = 1$ to $n_k$}
            \State Update $Z_i^k$ given $ Z_i^k, \theta_z^k $ and $A$
        \EndFor\vskip0.2cm
    \EndFor
        
    \StateX \textbf{(Outcome stage) } 
        \For{$m = 1$ to $M$}
        \State Set $R^k_{m i}=y^k_i - \sum_{q \neq m} g\left(\boldsymbol{x}^k_i ; \mathcal{T}_q, \mathcal{M}_q\right) - (\beta_z^{k'})^T \boldsymbol{Z}^k_i$ for $i=1, \ldots, n$ and $k=1,\ldots,K$.
        \State Propose a new tree $T_m^* \sim Q\left(\mathcal{T}_m \rightarrow \mathcal{T}_m^*\right)$.
        \State Compute the acceptance ratio
        \begin{align*}    
        r=\frac{\mathscr{L}\left(\mathcal{T}_m^*\right) Q\left(\mathcal{T}_m^* \rightarrow \mathcal{T}_m\right)}{\mathscr{L}\left(\mathcal{T}_m\right) Q\left(\mathcal{T} \rightarrow \mathcal{T}_m^*\right)},
        \end{align*}
        where $\mathscr{L}\left(\mathcal{T}_m\right)=p\left(\mathcal{T}_m \mid \boldsymbol{R}_m, \sigma\right) \times p\left(\mathcal{T}_m\right)$ and $\boldsymbol{R}_m=(R^1_{m1},\ldots,R^1_{mn_1},\ldots,R^k_{m1},\ldots,R^k_{mn_k})$.
        \State Set $\mathcal{T}_m = \mathcal{T}_m^*$ if $U \leq \min (1, r)$ where $U \sim \operatorname{Uniform}(0,1)$, otherwise retain $\mathcal{T}_m$.
        \State Sample the $\mu_{t \ell}$'s from their full conditional distributions.
        \EndFor    
        \State Sample $\tau'$ from its full conditional distribution.
        \State Sample $\beta_z^{k'}$ from its full conditional distribution.
        \StateX \textbf{(Monte Carlo estimation stage)} 
            \For{$g \in \mathcal{G}$}
            \State Predict $\hat{Y}^k_{i2}(g)$ given $Z^k_i$, $\boldsymbol{\mathcal{T}} = (\mathcal{T}_1,\ldots,\mathcal{T}_m),$ and $ \boldsymbol{\mathcal{M}} = (\mathcal{M}_1,\ldots,\mathcal{M}_m)$.
        \EndFor
        \State Average the potential outcomes over all units $\hat{\mu}(y,g) = \frac{1}{NK}\sum_{k=1}^K\sum_{i=1}^{n_k} \hat{Y}^k_{i2}(g)$
    \end{algorithmic}
\end{algorithm}

\section{Details of data generating process and estimators for the multiple network setting}\label{appendix:HLSM}

Given the generated multiple networks following HLSM in \eqref{eq:HLSM}, the individual behaviors at baseline are assigned as following 
\begin{align}\label{eq:y1}
    Y_{i1}^k = 5 + \beta_{u_1}U_{i1}^k + \beta_{u_2}U_{i2}^k + \epsilon_{i1}^k
\end{align}
where $\epsilon_{i1}^k \overset{iid}{\sim} N(0,1)$. We further specify a nonlinear outcome model for individual behaviors at follow-up as
\begin{align}\label{eq:y2}
Y_{i2}^k = 10 + 3Y_{i1}^k + 10 \text{sigmoid}(10 (G_i^k - 5)) + \beta'_{u_1} U_{i1}^k + \beta'_{u_2} U_{i2}^k + \epsilon_{i2}^k
\end{align}
where $\text{sigmoid}(x) = 1/(1+ e^{-x})$ and $\epsilon_{i2}^k \overset{iid}{\sim} N(0,1)$. Note that $G_i^k= \sum_{j \in \mathcal{N}^k_i} Y_{j1}^k / \sum_{j\in\mathcal{N}^k_i}A_{ij}$ where $\mathcal{N}^k_i$ is a set of neighbors that share a tie with node $i$ in the $k$th network. The parameter values are fixed at $\beta_\mathbf{u}=(\beta_{u_1},\beta_{u_2})=(0.5, 0.5)$ and $\beta'_\mathbf{u}=(\beta'_{u_1},\beta'_{u_2})=(10,10)$. With these true underlying values, the population average potential outcome is $\E(Y_{i2}(g)) = 25 +10 \text{sigmoid}(10 (g - 5))$.

In the estimation step to estimate latent locations, we set hyperparameters as $\sigma'_z = 5, \ \mu_{\alpha'}=0, \ \mu_{\beta'}=0,\  \sigma_{\alpha'}^2=100, \ \sigma_{\beta'}^2=100$ in \eqref{eq:HLSM}. Similar to the single network setting, three different estimators are compared, each using a distinct set of predictors in the BART model:

\begin{itemize}
    \item Oracle: $Y^k_{i2} = f(D_i^k) + \epsilon_i^k$ where $\mathcal{D}^k_i = (Y^k_{i1}, G^k_i, U^k_{i1}, U^k_{i2})$.
    \item LSM-semiBART: $Y^k_{i2} = f(D_i^k) + \beta_{z_1}^k Z_{i1}^k+ \beta_{z_2}^k Z_{i2}^k  + \epsilon_i^k$ where $\mathcal{D}^k_i= ( Y^k_{i1}, G^k_i)$.
    \item Naive: $Y^k_{i2} = f(D_i^k) + \epsilon_i^k$ where $\mathcal{D}^k_i =(Y^k_{i1}, G^k_i)$.
\end{itemize}

where $Z_{i1}^k$ and $Z_{i2}^k$ are estimated latent locations from HLSM in \eqref{eq:HLSM} and $f(\cdot)$ is modeled as the BART model.


\section{Supplementary plots for simulation and real data analysis}

 \begin{figure}[!t]
    \centering
    \includegraphics[width=1 \linewidth]{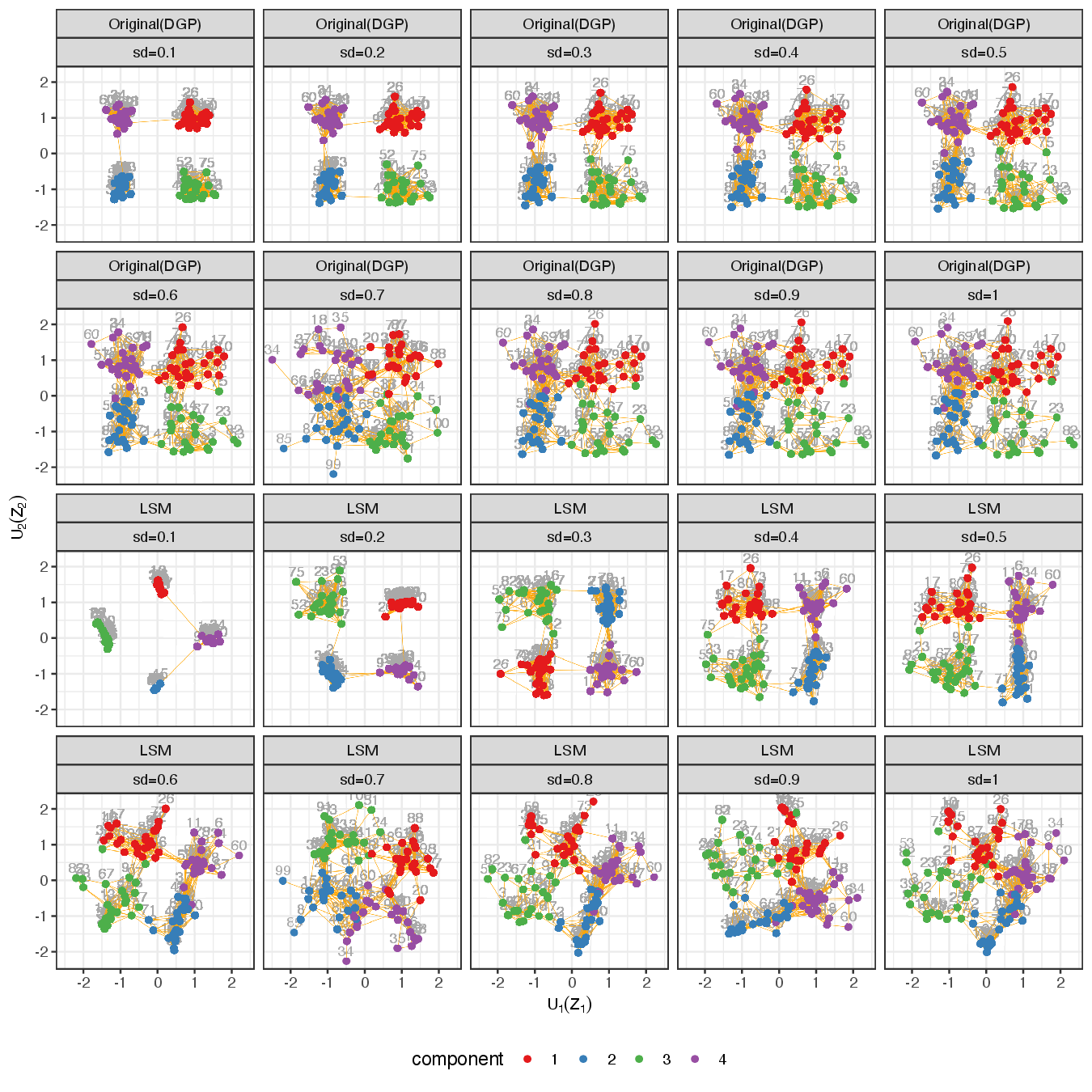}
    \label{fig:sim_sd}
        \caption{Simulation 1: The generated networks from the data generating process with $\sigma_j \in (0.1,\ldots,1)$ (top panels) and corresponding posterior mean of latent locations (bottom panels)}
\end{figure}



\begin{figure}[!h]
    \centering
    \includegraphics[width=0.9 \linewidth]{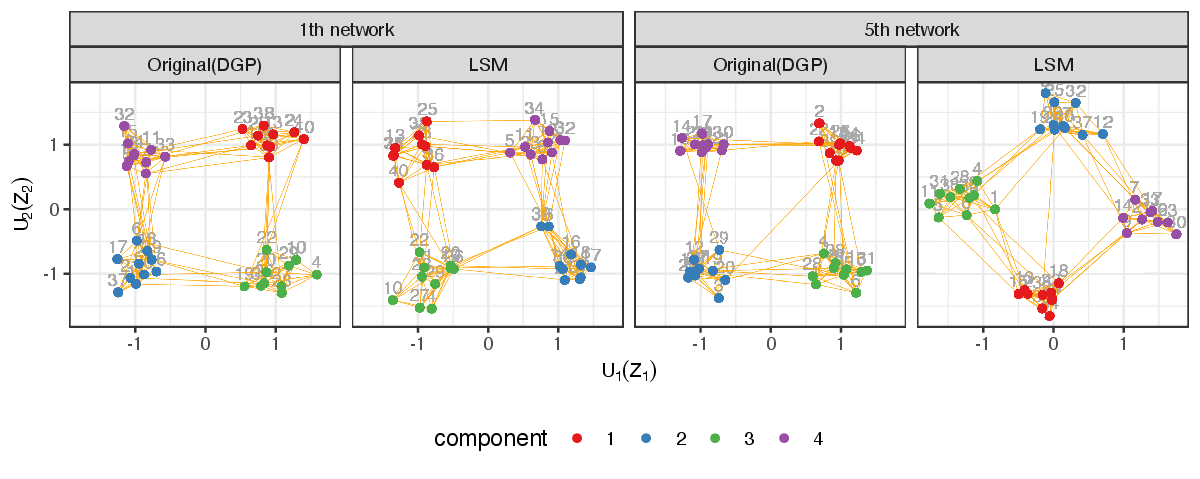}
    \caption{Simulation 3: The latent positions, $U_i$'s, from data generating process (left) and posterior mean of latent positions $Z_i$'s (right) corresponding to $1^{st}$ and $5^{th}$ network.}    
    \label{fig:HLSM_net}
\end{figure}

\begin{figure}[!t]
    \centering
    \includegraphics[width=1 \linewidth]{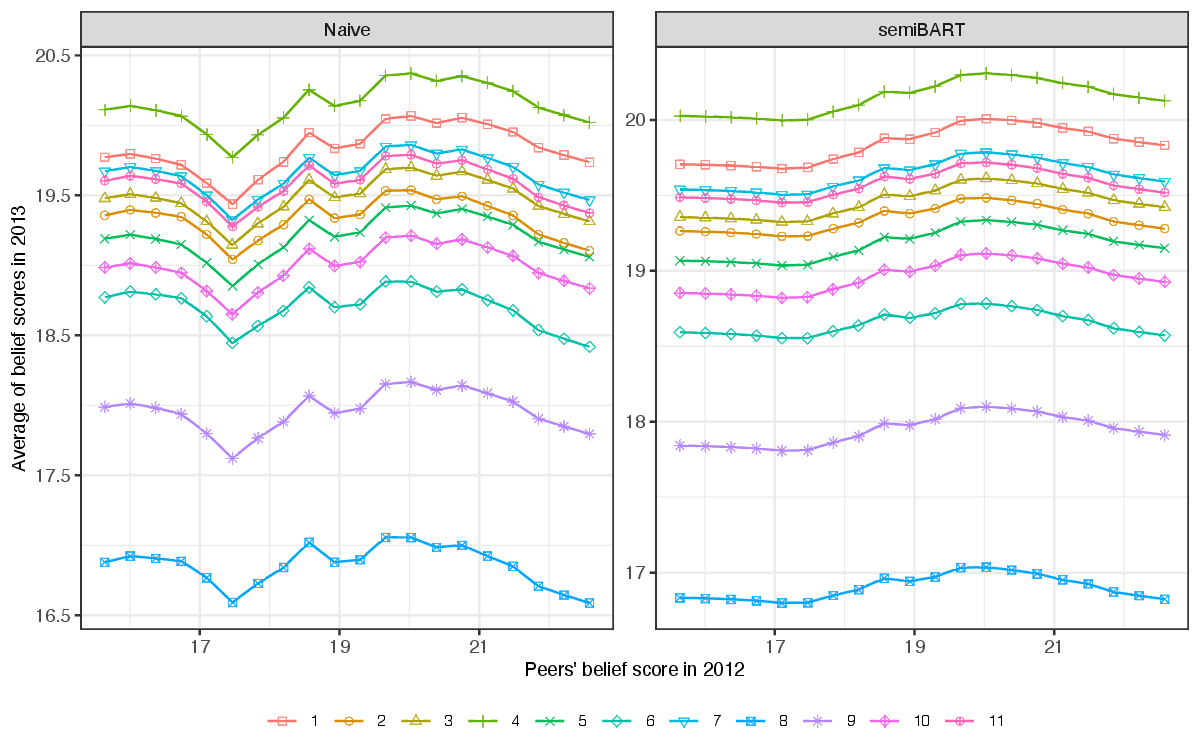}
    \caption{The estimated average belief scores in 2013 across all school from the Naive (left) and LSM-semiBART (right) estimator is represented for each school.}
    \label{fig:real_all}
\end{figure}

\newpage

\section{goodness of fit of HLSM}

In this section, we evaluate the goodness of fit of HLSM using advice-seeking network data. We perform posterior predictive checks by generating 11 networks from the parameters obtained at each MCMC iteration. Specifically, we utilize the latent space positions and the random intercepts $\alpha_0^k$ to generate these networks. We assess the model fit based on network density and transitivity, which can partially explain homophily. We use 200,000 MCMC iteration with 5000 burn-in thinning by every 200th step. 

\begin{figure}[!t]
    \centering
    \includegraphics[width=1 \linewidth]{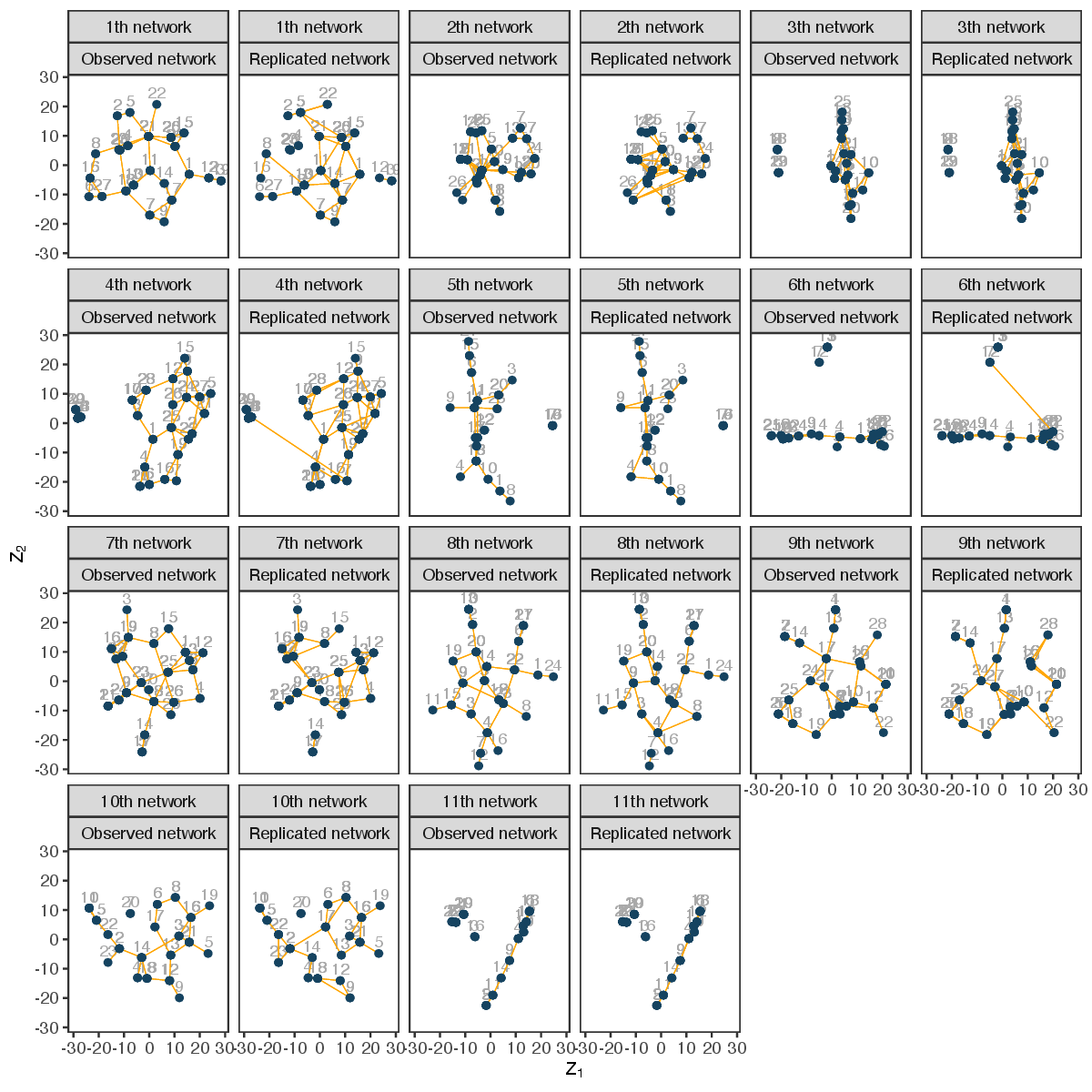}
    \caption{The illustration of observed and replicated network ties on estimated latent positions from a randomly selected MCMC sample.}    
    \label{fig:HLSM_fit_rep}
\end{figure}

First, we visualize an example of a replicated network alongside the observed network in Figure \ref{fig:HLSM_fit_rep}. The ties of both the observed and replicated networks are depicted on the estimated latent positions from a randomly selected MCMC iteration. Generally, this visualization demonstrates that the network ties from the replicated networks closely resemble those of the observed network.

\begin{figure}[!t]
    \centering
    \includegraphics[width=1 \linewidth]{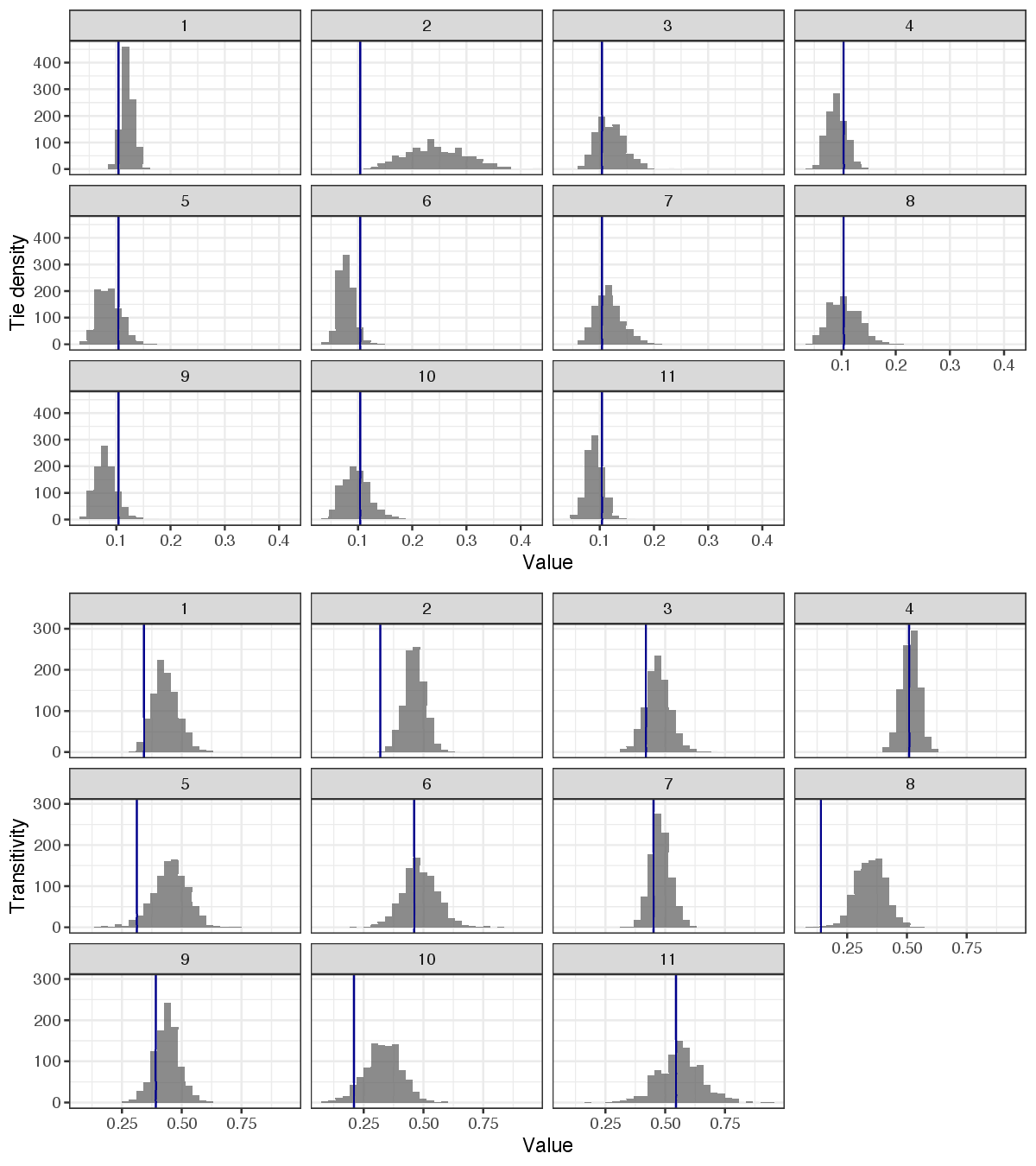}
    \caption{Posterior predictive checks of the HLSM: tie density (upper panel) and transitivity (lower panel).}    
    \label{fig:HLSM_fit_density}
\end{figure}

Figure \ref{fig:HLSM_fit_density} illustrates the distributions of network (tie) density and transitivity across the generated networks. The observed network statistic for each network are shown with a vertical line. In general, density and transitivity are adequately recovered indicating that the HLSM was able to capture these characteristics of the observed network. In School 2, however, both density and transitivity are overestimated potentially due to a high average degree (3.6), which results in closely positioned latent locations, thereby overestimating density and transitivity. To address this limitation and more effectively capture complex network characteristics, considering different dimensionalities of $Z$ for each school could be explored for future work.

\begin{figure}[!t]
    \centering
    \includegraphics[width=1 \linewidth]{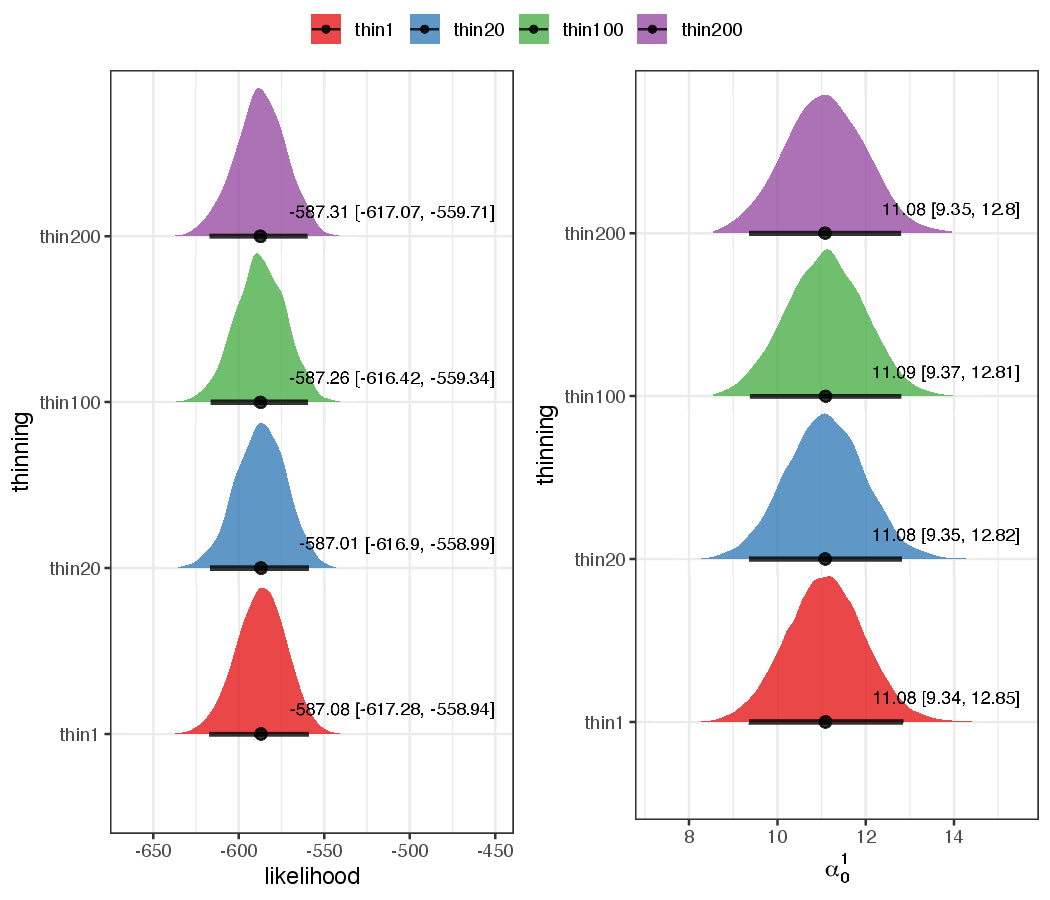}
    \caption{Density plot of the likelihood values computed at each MCMC iteration (left) and the posterior distributions of $\alpha_0^1$ (right) derived from advice-seeking networks for each thinning  (1,20,100,200). Each plot includes the posterior mean and the 95\% credible intervals, depicted on the bottom line and annotated within the plot.}    
    \label{fig:HLSM_thin}
\end{figure}

\end{document}